# Costs and benefits of phytoplankton motility

Peyman Fahimi,*[(a)] Andrew J. Irwin,*[(a)] Michael Lynch[(b)]

[(a)] *Department of Mathematics & Statistics, Dalhousie University, Halifax, NS B3H4R2 Canada.*
[(b)] *Center for Mechanisms of Evolution, Biodesign Institute, Arizona State University, Tempe, AZ 85287 USA.*
* *E-mail: fahimi@dal.ca; a.irwin@dal.ca*

## Abstract

The motility skills of phytoplankton have evolved and persisted over millions of years, primarily in response to factors such as nutrient and light availability, temperature and viscosity gradients, turbulence, and predation pressure. Phytoplankton motility is broadly categorized into swimming and buoyancy regulation. Despite studies in the literature exploring the motility costs and benefits of phytoplankton, there remains a gap in our integrative understanding of direct and indirect energy expenditures, starting from when an organism initiates movement due to any biophysical motive, to when the organism encounters intracellular and environmental challenges. Here we gather available pieces of this puzzle from literature in biology, physics, and oceanography to paint an overarching picture of our current knowledge. The characterization of sinking and rising behavior as passive motility has resulted in the concept of sinking and rising internal efficiency being overlooked. We define this efficiency based on any energy dissipation associated with processes of mass density adjustment, as exemplified in structures like vacuoles, gas vesicles, and frustules. We propose that sinking and rising are active motility processes involving non-visible mechanisms, as species demonstrate active and rapid strategies in response to turbulence, predation risk, and gradients of nutrients, light, temperature, and viscosity. In swimmers, internal and external motility efficiencies have been extensively explored in theoretical studies, yet they are largely neglected in energetic estimates within marine and freshwater sciences. Identifying the dissipative processes associated with swimming and buoyancy-regulating machinery offers deeper insight into motility costs relative to an organism's total metabolic rate.

## 1. Introduction

Phytoplankton are photosynthetic organisms, including some that are mixotrophic, which play pivotal roles in primary production and carbon cycling in aquatic ecosystems [1]. Investigating their locomotion - both swimming and sinking/rising - provides insights into marine and freshwater ecology, including biogeochemical cycling [2], dynamics of food webs [3], biological carbon pumps [4], population growth rates [5], and phenomena like marine snow and blooming [6]. Phytoplankton motility primarily enables vertical movement in the water column, allowing cells to access light near the surface and nutrients at depth. This repositioning ability can enhance growth and reproduction, potentially triggering bloom formation under favorable environmental conditions. Motile phytoplankton can also avoid nutrient-depleted layers and, in some cases, reduce exposure to grazers, thereby influencing population dynamics and trophic interactions.

Some species regulate their buoyancy or form sinking aggregates, facilitating the downward transport of organic matter and nutrients. This vertical export is a core process in the biological carbon pump, contributing to the sequestration of surface carbon in the deep ocean and supporting the formation of marine snow, aggregates that drive vertical fluxes of organic material and shape global biogeochemical cycles.

In this review, we delve into the costs and benefits associated with the locomotion of phytoplankton species. Specifically, we explore their swimming and buoyancy regulation over short distances in patchy aquatic environments and over long distances as part of migratory behaviors. Our examination encompasses both biological and physical dimensions of this phenomenon of oceanography shedding light on the gaps and potential future research directions.

Before evaluating the costs and benefits of phytoplankton motility, it is essential to identify which species are most significant in terms of both their abundance and their contribution to total primary production - an assessment that varies by geographical location. For instance, during the boreal winter of 2017 along a transect through the North Pacific Subtropical Gyre and the California Current System, the three most abundant groups were coccolithophores (35.5%), diatoms (25.2%), and dinoflagellates (19.5%) [7]. However, a global-scale dataset spanning 2002–2022 indicates a different abundance ranking: chlorophytes (e.g., green algae) (~26%) are most prevalent, followed by diatoms (~24%), haptophytes (e.g., coccolithophores) (~15%), cryptophytes (~10%), cyanobacteria (~8%), and dinoflagellates (~3%) [8]. When considering primary production globally (from 1998 to 2011), diatoms contribute about 50% of total phytoplankton production, while coccolithophores and chlorophytes each contribute roughly 20%, and cyanobacteria around 10% [9].

It is also important to note several key distinctions between marine and freshwater systems. First, the dominant phytoplankton groups differ between marine and freshwater environments. Marine communities are typically composed of diatoms, coccolithophores, green algae, dinoflagellates, and silicoflagellates, with cyanobacteria occasionally abundant, whereas freshwater systems are more often dominated by cyanobacteria, green algae, chrysophytes (golden-brown algae), and dinoflagellates, depending on the nutrient status of the environment [10]. Second, periodic vertical migration is generally more prevalent in marine environments than in freshwater systems, reflecting fundamental constraints on buoyancy regulation in freshwater, including the lower density of the medium and the absence of very large phytoplankton cells [11]. Nevertheless, vertical migration in freshwater systems is well-documented for cyanobacteria, which regulate their position in the water column through buoyancy control mediated by gas vesicles and carbohydrate ballast [12].

Swimming and floating in water might be energetically costly for microorganisms at least for two reasons:

(A) Their small size and slow speeds place them in the low-Reynolds-number regime: In this regime, inertial forces are negligible compared to dominant viscous forces, causing water to behave like a thick, sticky fluid - more like honey than a free-flowing liquid. As a result, even the smallest movement requires a constant energy input; if the organism stops pushing, it stops almost immediately. For example, a tiny microorganism moving at about 30 microns per second will barely continue moving if the propelling force suddenly stops - traveling only about 0.1 angstrom before coming to a halt. This deceleration occurs in just 0.6 microseconds [13].

(B) Their motility machinery may be highly inefficient: In microswimmers, the mechanochemical efficiency of swimming machinery has been extensively investigated by biophysicists but has been largely neglected by studies in marine and freshwater sciences, leading to systematic underestimation of the energetic costs of motility. Similarly, in buoyancy-regulating organisms, although the molecular mechanisms underlying buoyancy control have been studied in detail by marine and freshwater scientists, the efficiency of buoyancy-regulating machinery has been almost entirely overlooked by both biophysicists and oceanographers, again resulting in underestimation of the energetic costs of buoyancy regulation.

There are notable differences between swimming and sinking/rising, resulting in varying cost-benefit analyses across species. For example, diatoms demonstrate sinking and rising behaviors, whereas dinoflagellates possess the capability for both swimming and buoyancy regulation. This differentiation might address certain unresolved issues in the literature concerning variations in population growth rates among different phyla. For example, why do dinoflagellates typically exhibit lower population growth rates compared to diatoms of similar size, despite having higher protein, chlorophyll, and carbohydrate content [14,15]? The deeper question is why an organism would want to do this, as selection puts a primacy on offspring production. Presumably, what is lost in growth rate is compensated by higher survival, i.e., avoidance of predators?

It has been suggested that dinoflagellates coexist with other phytoplankton species by using their swimming abilities to counteract physiological disadvantages [16]. However, does the benefit of swimming always outweigh its costs? In this study, through an extensive analysis of swimming costs, we demonstrate that these costs may be remarkably high relative to the total metabolic rate of the organism, even under conditions favorable for active growth. The significant contribution of such high metabolic costs may compel the organism to allocate less energy to growth, resulting in relatively lower population growth rates.

This review follows a logical framework: to estimate motility costs and benefits, one must consider four key components (Figure 1):

1- **Construction Costs:** The first step is to examine the energetic costs of building the motility machinery, which includes both intracellular components and extracellular, surface-related structures - hereafter referred to as internal and external machinery, respectively. These construction costs can be either one-time investments or ongoing costs per unit time. They are typically quantified in terms of ATP molecules (the energy currency of living organisms) or standard energy units such as joules. To evaluate their significance, these costs can be compared to the organism's total chemical energy investment and turnover, including its protein, lipid, and carbohydrate reserves, yielding a percentage estimate of investment in motility.

2- **Operational Costs:** The second component involves assessing the energy required to operate the internal and external motility machinery. This is more complex than the construction cost, as it must account not only for the ATP demands of molecular processes driving motility but also for energy losses due to dissipation (i.e., heat produced by inefficient systems). These operational costs per unit time should be compared to the organism's total metabolic rate - specifically, its total ATP production rate, which in phytoplankton primarily arises from mitochondrial and chloroplast activity under aerobic conditions.
3- **Environmental Modulation:** One must examine how environmental parameters - such as temperature, light, nutrient availability, viscosity, turbulence, viruses, and predators - affect motility costs, both directly and indirectly. This is arguably the most complex aspect of the problem, as environmental influences are multifaceted and difficult to quantify precisely.
4- **Benefits:** Assessing the benefits of motility primarily involves determining how - and to what extent - it enhances nutrient uptake, access to optimal light conditions, and predator avoidance. However, these benefits are strongly influenced by environmental conditions and geographical location, which shape nutrient and light availability as well as predation pressure.

The dissipative costs of motility machinery determine motility efficiency, which can be divided into external and internal efficiency. Motility external efficiency is influenced by cell surface structures, including the geometric shape of the cell body [17,18] and the external features of motility-related structures, such as flagella (e.g., their length, number, geometry, angle, beat frequency, beat amplitude, and arrangement) [13,19–21]. In contrast, motility internal efficiency is governed by dissipative processes within intracellular structures, such as the axoneme [22,23], molecular motors, type IV pili, frustule, vacuole, gas vesicle, and coccolith.

Because swimming and buoyancy regulation arise from distinct intracellular mechanisms, their associated costs also differ. Swimmers, such as flagellated phytoplankton, invest in constructing and operating flagella [24], whereas in sinkers like diatoms, differences in mass density between cellular components (e.g., vacuoles) and structural compartments such as the frustule play significant roles in buoyancy regulation. This regulation is achieved through processes such as substituting heavy ions with lighter ones (and vice versa) within vacuoles, or by modifying the frustule through changes in composition (e.g., silica precipitation and Si:C ratio) and structure (e.g., frustule volume and cytoskeletal organization of silica-depositing vesicles) [25–28].

Because buoyancy regulating species are frequently likened to passive particles in the literature, the assessment of energy dissipation linked to their intracellular machineries has been neglected. Our perspective is that sinking and rising are active mechanisms for motility, using non-visible machinery. In this paper, we discuss the biological mechanisms involved in sinking internal efficiency, while the underlying physics needs to be formulated in future studies. The emphasis on non-visible motility mechanisms stems from the fact that determining the structure and function of visible motility machinery (such as flagella) is relatively easier with video microscopy, making it more accessible for researchers to investigate its efficiency. In contrast, studying non-visible motility mechanisms, such as those mediated by buoyancy-regulating systems like the frustule, vacuole, gas vesicle, and coccolith - found in diatoms, cyanobacteria, coccolithophores, and others - is more challenging.

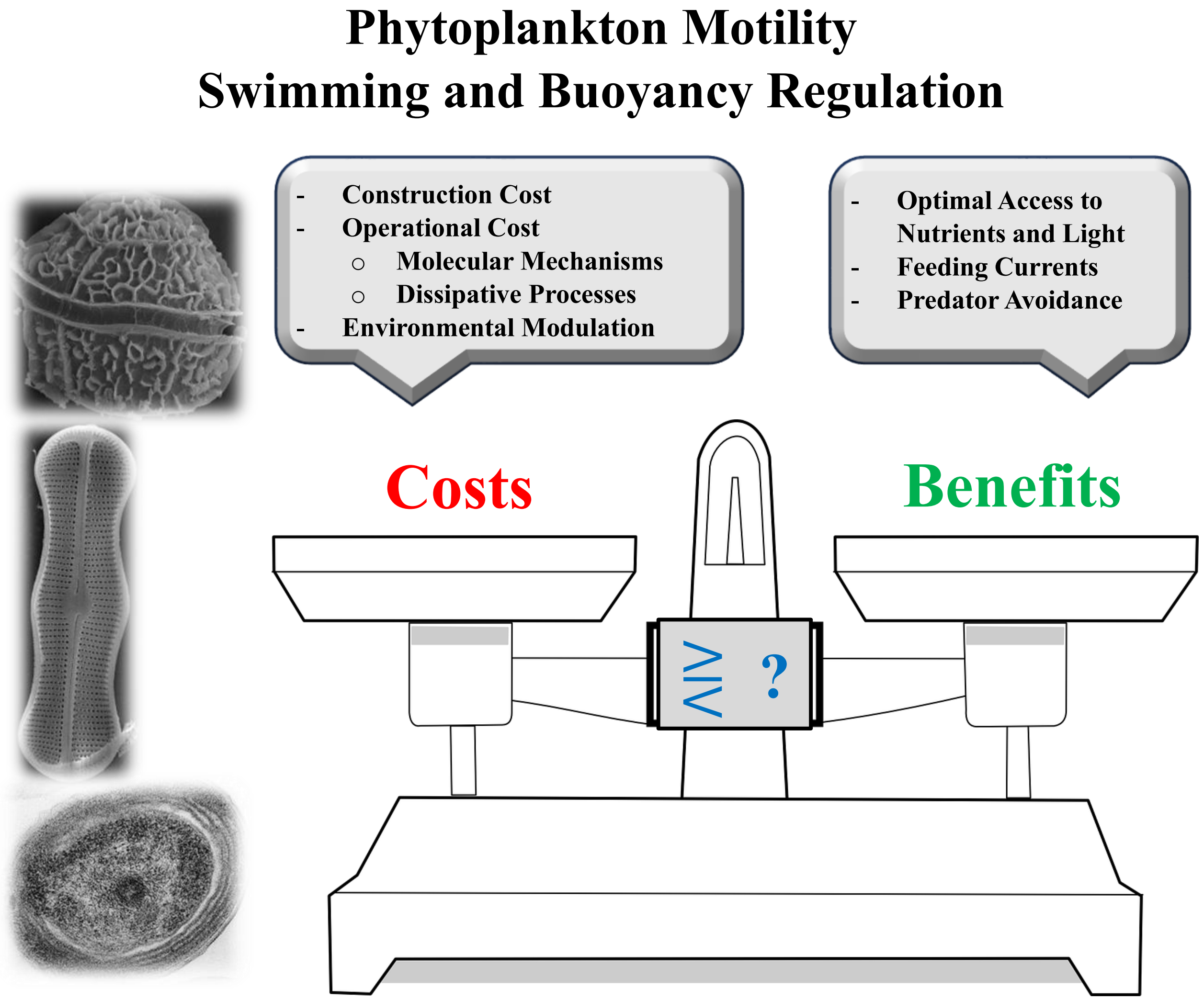

*Figure 1: A conceptual depiction of the trade-off between the costs and benefits of phytoplankton motility in both swimmers and buoyancy regulators. The major components contributing to this trade-off are highlighted in two boxes. On the left, representative phytoplankton species are shown from top to bottom: a dinoflagellate, a diatom, and a cyanobacterium. Images are sourced from the public domain via Wikipedia.*

The following sections build the framework of this review, beginning with the fundamentals of swimming and buoyancy regulation (Sections 2–3), followed by their environmental context through migration (Section 4), and culminating in a cost–benefit analysis of motility (Section 5). We then examine how environmental drivers - particularly temperature, along with climate change and acidification - modulate these costs and benefits (Section 6). Readers are encouraged to carefully examine Figure 2 before proceeding, as it provides a structured overview of the narrative and organization of the review. To further facilitate navigation between sections, Table 1 summarizes the correspondence between each motility mode, its mechanistic description, and its associated energetic analysis.

- **Section 2 Outlines the Fundamentals of Phytoplankton Swimming Machinery**
  - Structure and function of swimming machinery (flagellated phytoplankton, cyanobacteria, and escape-jumping phytoplankton).
  - Stokes-law constraint as a lower bound on the operational cost of swimming.
  - Efficiency of swimming machinery as a major contributor to operational cost.
  - Swimming in viscosity gradients shaped by environmental and biological processes.
  - Functional benefits of swimming (taxis, feeding currents, nutrient uptake, predator avoidance).

- **Section 3 Outlines the Fundamentals of Phytoplankton Buoyancy Regulation Machinery**
  - Buoyancy-driven Stokes velocity expression as a baseline for identifying physical drivers and underlying molecular mechanisms of buoyancy regulation.
  - Mass density of cells and intracellular components.
  - Molecular mechanisms of density modulation in cyanobacteria, coccolithophores, and diatoms.
  - Multifaceted role of lipids in buoyancy regulation.
  - Modified Stokes-based formulations accounting for intracellular density heterogeneity.
  - Efficiency of buoyancy regulation machinery as a largely overlooked concept.

- **Section 4 Outlines the Fundamentals of Phytoplankton Migration**
  - Migration-driven exposure to variable environmental conditions and associated changes in motility costs and benefits.
  - Environmental gradients encountered during migration (temperature, light, nutrients, viscosity, predation/viral risk, and turbulence).
  - Roles of fast vs. slow migrators, migration distance, and vertical distribution/layer formation for optimal light and nutrient access.
  - Adaptive strategies in heterogeneous/turbulent environments (e.g. chain formation, subpopulation division) and their associated costs and benefits.

- **Section 5 Outlines Cost–Benefit Analysis of Motility**

  **Benefits of Motility**
  - Context-dependent benefits (chemotaxis, phototaxis, feeding currents, nutrient uptake, predator avoidance, jump-like movements).
  - Sherwood number as a measure of transport enhancement.
  - Growth rate of motile vs. non-motile cells.

  **Costs of Motility**
  - Construction costs of swimming machinery in flagellated phytoplankton and cyanobacteria (synthesis and remodeling).
  - Construction costs of buoyancy regulation machinery in cyanobacteria and aflagellated phytoplankton (coccolithophores and diatoms).
  - Efficiency of swimming machinery (flagellated phytoplankton and cyanobacteria).
  - Efficiency of buoyancy regulation machinery (cyanobacteria, coccolithophores, and diatoms).
  - Direct and indirect motility costs linked to predator and viral risk.
  - Case study: construction and operational costs of swimming in dinoflagellates relative to total metabolic rate.

- **Section 6 Outlines the Effects of Temperature on Motility Costs, with Additional Roles of Climate Change and Acidification**
  - Temperature and acidification effects on phytoplankton motility.
  - Temperature-driven changes in viscosity, density, organism size, shape, swimming speed, flagellar assembly and function, and ATP production/consumption rate.

*Figure 2: Schematic overview of the structure and content of each section of this review.*

*Table 1: Cross-referencing of motility modes with their mechanistic descriptions and energetic analyses across the review.*

| Organism(s) | Motility Mechanism | Construction Cost | Operational Cost |
|---|---|---|---|
| Flagellated Phytoplankton | Section 2.1 | Section 5.2.1 | Section 5.2.6 |
| Swimming Cyanobacteria | Section 2.2 | Section 5.2.2 | Section 5.2.7 |
| Escape-jumping Phytoplankton | Section 2.3 | --- | Section 5.2.8 |
| Buoyancy-Regulating Cyanobacteria | Section 3.2 | Section 5.2.4 | Section 5.2.10 |
| Coccolithophores | Section 3.3 | Section 5.2.5 | Section 5.2.11 |
| Diatoms | Section 3.4 | Section 5.2.5 | Section 5.2.12 |

## 2. Mechanisms and Dynamics of Phytoplankton Swimming

To estimate the construction and operational costs of phytoplankton swimming, one must first understand the various mechanisms by which these organisms achieve motility. These mechanisms include: the use of flagella, as seen in dinoflagellates, chrysophytes, raphidophytes, haptophytes, and chlorophytes; the generation of surface waves powered by protein motors in some cyanobacteria; gliding via type IV pili in certain cyanobacteria; and specialized structures such as haptonema used for escape jumps. In addition to understanding the molecular basis of these motility machineries and their associated energetic demands, it is also essential to consider the efficiency of the swimming machinery involved. These two factors contribute to the operational cost of swimming. A complementary question is how swimming benefits the organism by generating feeding currents and enhancing nutrient uptake rates? To address this question, one must analyze the fluid velocity fields generated by flagella or other swimming mechanisms.

### *2.1. Flagellated Phytoplankton*

Most phytoplankton swimmers use flagella, which are long, whip-like structures common in eukaryotic cells that facilitate movement. These flagella typically consist of an internal structure called the axoneme, composed of a "9+2" arrangement of microtubules: nine outer doublets surrounding two central singlets [24]. Protein complexes like dynein arms, nexin-DRC, and radial spokes link these microtubules, all enclosed within a membrane [24]. Flagellate swimming speeds typically range from 10 to 1500 μm/s, with a median around 130 μm/s, based on data compiled by Ref. [29]. The speed of flagellated phytoplankton and the flow fields they generate are influenced by the number of active flagella [24], flagellum length [24], beat frequency, beat amplitude, beat synchronization [30], and flagellar arrangement [31]. Data on the number and length of flagella have been compiled for over 200 species of unicellular eukaryotes, including phytoplankton, and bacteria [24]. As an example of the flagellar arrangement in phytoplankton species, dinoflagellates possess two flagella, one transversal and the other longitudinal (known as the trailing flagellum). Fenchel [32] observed that the trailing flagellum contributes to both the cell's translational velocity and its rotation around an axis that is perpendicular to its length. Meanwhile, the transversal flagellum mainly induces rotation around the cell's longitudinal axis.

## 2.2. Swimming Cyanobacteria

There are alternative forms of swimming, other than employing flagella in phytoplankton, which are not easily observable. In *Synechococcus*, locomotion is facilitated by traveling waves along the surface of the cell body [33]. Given its small size (≈0.8–1.5 μm), rotational Brownian motion may partially randomize swimming direction, potentially limiting sustained directional (taxis-driven) motion [34]. The surface waves result from cyclic surface deformations driven by protein motors modeled as a rotating helix. Motors move faster than waves traveling on the body surface, and the waves move faster than the organism's swimming speed. Consider a cell with a radius $r$ of 1 μm, with surface waves characterized by a dimensionless parameter $\epsilon = 0.02$ that determines the relative deformation of the radius (i.e., a deformation amplitude $\epsilon r = 0.02$ μm), and a dimensionless wave number $k = \frac{2\pi r}{\lambda}$, where $\lambda$ represents the wavelength. To achieve an organism velocity $v$ of 15 μm/s, a wave speed $c = \frac{\omega r}{k}$ of 456 μm/s is necessary, where $\omega$ represents the angular frequency, as determined by the formula $v = 82.2\epsilon^2 c$ derived in Ref. [33]. The fixed dimensionless coefficient 82.2 in this equation arises from an approximate representation of cell-surface deformations as Legendre polynomials and a fixed wave number $k$=16. For relatively large wave numbers, the result will change. For a complete discussion, see Ref. [33] and exploration of other models investigating the swimming velocity driven by surface-distorted waves, see [33,35,36]. The organism's speed contributes to the Stokes hydrodynamic swimming cost, while the wave speed on the surface of the cell body and the speed of motor proteins contribute to energy dissipation in the internal machinery.

The freshwater cyanobacterium, *Synechocystis* sp., exhibits a gliding or twitching-type motility with an average speed ranging from 0.09 to 0.15 μm/s [37]. This movement in water is facilitated by type IV pili (T4P), which are long, hair-like dynamic filaments on the organism's surface [38,39]. The molecular structure of T4P is detailed in [40]. Although the physics governing the velocity field patterns, thrust forces, and swimming efficiency of this cyanobacterium remain unexplored, it is suggested that its motility resembles that of soil bacteria *Myxococcus xanthus* [41]. The motility of *Myxococcus xanthus*, viewed as an elastically flexible cell, can be modeled by several forces: the forces generated by motor proteins interconnected by angular springs, viscous forces, and interaction forces [42].

## 2.3. Escape-Jumping Phytoplankton

High-speed escape jumps, which are purposefully directed, have been observed in various phytoplankton swimmers, demonstrating their ability to evade predators by responding to fluid mechanical cues in the surrounding water [43–45]. For instance, *Chrysochromulina simplex* has been shown to increase its swimming speed from approximately 1,400 μm/s to 6,600 μm/s (a nearly fivefold jump) in just 0.05 seconds [43]. However, jumping is not only a predator avoidance behavior. In the photosynthetic ciliate *Mesodinium rubrum*, spontaneous jumping serves as a key behavioral adaptation to overcome diffusion-limited nutrient uptake [46]. The characteristics of these jumps vary with temperature, strain, and light conditions. Despite differing speeds and environments, both Antarctic and temperate strains of *M. rubrum* achieve similar jump distances [46]. The Antarctic strain compensates for cold temperatures and reduced metabolic rates by performing longer-duration jumps with extended ciliary beat cycles, whereas the temperate strain exhibits short-distance tumbling, which may further enhance nutrient uptake by generating local shear [46]. The molecular mechanisms underlying escape jumps in microswimmers vary across species. In haptophytes such as *Chrysochromulina simplex*, the escape jump is a coordinated two-step process [45]. First, the haptonema (a long, flexible appendage) rapidly coils, briefly pulling the cell forward. Immediately afterward, the two flagella reverse their beating direction and dramatically increase their speed, propelling the cell backward at high speed [45]. The energetic cost of these high-speed jumps will be addressed in Section 5.

### *2.4. Stokes' Law*

Phytoplankton swimmers operate in a low Reynolds number regime, where the dimensionless Reynolds number $\mathrm{Re} = \frac{rv\rho}{\eta}$ is much less than one. In this regime, inertial forces are negligible compared to viscous forces, fundamentally shaping how these organisms move. Here, $r$ and $v$ denote the organism's characteristic size and swimming speed, while $\rho$ and $\eta$ represent the fluid's density and dynamic viscosity, respectively.

At low Reynolds numbers, the nature of the forces acting on microorganisms fundamentally determines the structure of the surrounding flow field. A fundamental physical distinction exists between swimming and sinking/rising microorganisms in low-Reynolds number flows. Actively swimming cells are, to leading order, force-free systems in which propulsion balances viscous drag, generating a dipolar flow field described by a stresslet [47]. In contrast, sinking or rising cells experience a net external force due to gravity (or buoyancy), producing a monopolar flow field described by a Stokeslet [47].

The Navier-Stokes equation, which governs motility in a viscous fluid, describes the momentum balance in a Newtonian fluid based on Newton's second law. For Newtonian fluids, the viscosity remains constant, meaning the fluid's resistance to flow does not change with the speed at which it flows or the amount of force applied. Under these conditions, the drag force, $F$, acting on a spherical, rigid particle in laminar and uniform flow - assuming non-interfering particles - is given by $F = 6\pi\eta rv$, where $6\pi\eta r$ is the drag coefficient and $\pi \cong 3.14$ is a constant. The minimum power required for swimming by a spherically shaped species is then:

$$P = Fv = 6\pi\eta rv^2 \qquad (1).$$

For a thorough derivation, readers are directed to fluid mechanics textbooks such as [48], specifically the section titled 'Creeping Flow around a Sphere.' For detailed, biologically relevant discussions, see 'Foundations 16.2' in Ref. [49]. Equation (1) has found extensive use in the literature for estimating the swimming cost of microswimmers, however, this equation represents the minimum power needed, not the total power required for swimming. Consequently, many marine and freshwater studies have incorrectly relied exclusively on this equation, portraying the swimming cost of phytoplankton as negligible relative to total metabolism and substantially underestimating the true energetic cost of swimming.

### *2.5. Swimming Efficiency*

Mechanical efficiency is defined as the ratio between the minimum power needed to tow an object (characterized by its drag coefficient at the swimmer's average speed over a beating cycle), and the average power the swimmer transfers to the surrounding fluid during that same cycle [47,50]. Several methods exist for determining the denominator of this ratio, though the underlying physics is beyond the scope of this review. For instance, calculating the local velocities and corresponding forces along the length of the organism body and flagella (or other locomotion structures) typically involves applying theories such as resistive force theory or slender-body theory [51,52]. These approaches require evaluating the strain rate tensor to capture the interactions with the surrounding fluid [52]. Therefore, for the purposes of this review, we rely directly on empirical values of swimming efficiency reported in the literature to estimate the operational cost of swimming, as detailed in Section 5.

### *2.6. Viscosity Gradients*

Looking at Equation (1), it's important to note that viscosity is not necessarily constant in the ocean. Phytoplankton motility may take place within a viscosity gradient. Viscosity is influenced by water properties and by the biological activity of microorganisms within shear-induced aggregates or associated with extracellular polymeric substances (EPS) of high molecular weight, as well as light-induced lysis [53]. Recent microrheological techniques [53] have revealed that approximately 45% of the studied phytoplankton species (including cultures of diatoms, dinoflagellates, and haptophytes) experience viscosity levels up to 40 times that of artificial seawater within a 30 μm radius from the cell (phycosphere). According to [53], the relative viscosity is, on average, 2.6 times higher near the cell wall due to the production of extracellular polymeric substances, up to 30% higher near lysed and dead cells, and more than one order of magnitude higher within aggregates. How do these viscosity gradients affect the motility of phytoplankton, and do they employ any strategies in response? Microfluidic experiments involving *Chlamydomonas reinhardtii* by Stehnach *et al.* [54] indicate that these organisms encounter high local viscosities, which decrease their speed. Nevertheless, they exhibit adaptability by changing their strategy and reorienting down the gradient [54].

### *2.7. Swimming Benefits*

Different forms of taxis in phytoplankton, including chemotaxis, phototaxis, and gravitaxis, enable organisms to efficiently increase encounter rates with nutrient-rich patches and access light over optimal timescales, representing a central functional benefit of motility. Predator avoidance constitutes another important functional advantage under ecological pressure. However, the relative importance of these benefits is context-dependent. In contrast, enhanced nutrient uptake through feeding currents generated by motility machinery represents a directly quantifiable energetic benefit.

Phytoplankton can use their motility apparatus to generate feeding currents in the surrounding fluid, thereby enhancing the likelihood of prey or particle capture and facilitating greater nutrient uptake. The resulting flow velocity fields - driven by forces associated with different flagellar configurations - determine the shape and extent of streamlines around the organism [31] and can be modeled using idealized force-based approaches (see Appendix Table S1). These point-force models include the *stokeslet* [55], *puller stresslet* [55,56], and *three-stokeslet quadrupole* [57]. They are formulated based on the action of one or more time-averaged point forces applied at reference points, with flow velocity fields derived from Green's function solutions [58]. In these models, the propulsive force $F$ is considered proportional to the product of flagellum length, number of active flagella, beat frequency, and beat amplitude [31]. The modeled flow velocity fields can be used to estimate clearance rates - that is, the volume of fluid per unit time from which prey or particles can be captured. For quantitative estimates of clearance rates and numerical simulations of flow fields, the reader is referred to [31]. While this approach offers a mechanistic assessment of clearance rates generated by swimming, we solely rely on empirical equations in Section 5 to estimate the enhancement in nutrient uptake rate as a measure of motility benefit.

## 3. Essential Principles of Buoyancy Regulation

In this section, we discuss the Stokes-driven buoyancy equation, focusing on the parameters it captures completely, partially, or not at all. We explore the developed form of this equation, emphasizing the differences in mass density among various organelles of the species, which can deviate from the mean mass density of the whole organism. The foundational knowledge presented in this section is essential for understanding and quantifying the cost of buoyancy regulation, encompassing both its molecular mechanisms and associated dissipative processes.

### *3.1. Buoyancy-Driven Stokes Velocity*

When a spherical rigid object falls through a fluid, three forces act upon it:

1) Gravitational Force: This force pulls the object downward and is determined by the object's mass, $F_{\text{gravity}} = mg$, where $m$ is the object's mass and $g$ is the gravitational acceleration (standard value $g = 9.80665$ m/s²). For a sphere, the mass is given by $m = \rho_c V$, where $\rho_c$ is the mass density of the sphere and $V = \frac{4}{3}\pi r^3$ is its volume. Substituting $m$, we get $F_{\text{gravity}} = \frac{4}{3}\pi r^3 \rho_c g$.
2) Buoyant Force: This upward force arises due to the displaced fluid and is expressed as $F_{\text{buoyant}} = \rho_w V g$, where $\rho_w$ is the mass density of the fluid.
3) Drag Force: This resistive force opposes the object's motion through the fluid. According to Stokes' law, $F_{\text{drag}} = 6\pi\eta r v$, where $\eta$ is the dynamic viscosity of the fluid, $r$ is the radius of the sphere, and $v$ is its velocity. By balancing these forces, $F_{\text{gravity}} - F_{\text{buoyant}} = F_{\text{drag}}$. Substituting the expressions for each force and solving for $v$, the sinking/rising velocity is:

$$v = \frac{2gr^2(\rho_c - \rho_w)}{9\eta} \qquad (2),$$

where $\rho_c - \rho_w$ represents the difference between the mass density of the object (e.g., a sinking or rising phytoplankton) and the fluid (e.g., seawater or freshwater). While treating sinking phytoplankton as passive particles implies that sinking, regardless of speed, incurs no direct energetic cost to the organism (because the organism does not pay for the dissipation - gravity does the work), this perspective overlooks the reality that many species actively regulate their buoyancy (and thus their sinking or rising speed). When viewed as active organisms, their motility costs can be meaningfully estimated based on the metabolic energy required for buoyancy control mechanisms, which based on Equation (2), depend at least on density adjustments and shape adaptations. In this section, we discuss the factors that actively influence sinking and rising speeds, which directly impact the associated energetic costs.

Phytoplankton exhibit varying density differentials relative to their environment. The mass density of phytoplankton typically ranges from 920 to 1085 kg $m^{-3}$ for cyanobacteria, 1020 to 1140 kg $m^{-3}$ for Chlorophyta, and 1009 to 1263 kg $m^{-3}$ for Bacillariophyta [59–61]. Intracellular macromolecules such as proteins and carbohydrates generally have higher mass densities (~1300 and ~1500 kg $m^{-3}$, respectively) than seawater (~1030 kg $m^{-3}$), whereas lipids have a lower mass density (~860 kg $m^{-3}$) compared to water [62]. Phytoplankton employ several mechanisms to adjust their internal density, enabling vertical migration (both upward and downward) over timescales ranging from a day to several days or even up to a week [63].

### *3.2. Buoyancy Regulating Cyanobacteria*

In cyanobacteria, buoyancy regulation can occur through both changes in gas-vesicle state (e.g., synthesis or collapse by turgor pressure) and changes in intracellular ballast materials such as carbohydrates [12]. In species such as *Microcystis aeruginosa*, where gas vesicles are too strong to collapse, buoyancy changes are achieved by light-driven accumulation of carbohydrates during photosynthesis (increasing cell mass density and causing sinking) and subsequent consumption in the dark during respiration (decreasing mass density and causing rising) [64–66]. In *Microcystis*, environmental conditions further modulate this process [65]: higher temperatures increase the rates of photosynthetic carbohydrate storage during the light period and its respiratory depletion in the dark, leading to more rapid changes in cell mass density and shorter migration cycles; increasing water turbidity reduces light penetration and confines vertical migration to shallower depths; and increasing colony size increases migration velocity and frequency, promoting more regular vertical migration patterns [65].

### *3.3. Coccolithophores*

Eukaryotic phytoplankton employ multiple mechanisms to regulate their sinking and rising behavior, including ion substitution (replacing lighter ions with heavier ones, or vice versa) [25], accumulation of polysaccharides [67], cell wall synthesis [67], active transport of water from the cytosol into the central vacuole [68], and incorporation of low-density compounds such as trimethyl ammonium (~993 kg $m^{-3}$) into the vacuole [62].

In coccolithophore *Emiliania huxleyi*, mass density (and thus sinking behavior) is modulated by three main intracellular mechanisms under nutrient limitation, all acting through changes in cellular composition. First, calcification is the dominant driver: coccoliths composed of dense calcite can account for ~64–73% of total cell mass, so increased calcification substantially raises cellular density and sinking rate [67]. Second, carbohydrate accumulation (e.g., storage polysaccharides) increases cell mass more modestly, leading to a smaller but still measurable increase in density and sinking [67]. Third, osmolyte substitution, specifically replacing lighter compounds (e.g., $TMA^+$, TMG, $Cl^-$) with the denser molecule dimethylsulfoniopropionate (DMSP), has only a negligible effect on density because organic solutes represent a small fraction of total cell mass [67].

### 3.4. Diatoms

Density variations in diatoms arise from a complex interplay among factors such as the heavy siliceous frustule (denser than 2000 kg $m^{-3}$), ammonium derivatives, cell shape, nutrient flux, silica requirements, and large vacuoles which are influenced by environmental conditions and grazing pressure [28,62]. The silicification process, which defines the frustule, significantly influences vertical migration of diatoms when high energy resources are not available [27,69]. When faced with nutrient limitations and iron deficiencies, diatoms tend to exhibit higher Si:C ratios, promoting faster sinking rates to access nutrient-rich deep ocean layers [28].

Are diatoms limited to sinking, or can they regulate their buoyancy to become neutral or positively buoyant? Numerous studies have demonstrated that diatoms are capable of achieving neutral or even positive buoyancy. For instance, laboratory experiments reveal that *Thalassiosira weissflogii* maintains neutral buoyancy under nutrient-rich conditions but begins to sink when nutrients are scarce [70]. Similarly, in *Rhizosolenia formosa H. Peragallo*, nitrate depletion reduced the proportion of positively buoyant cells from 11% to 4%, which rebounded to 9% within 12 hours of nitrate resupply [71]. Research on the genus *Rhizosolenia* also indicates that buoyancy direction is influenced by light intensity, likely through a mechanism involving carbohydrate ballasting [72]. Field studies further support this view: prior to the spring bloom, non-growing diatoms tend to sink, whereas during the bloom, they exhibit neutral buoyancy, suggesting that buoyancy regulation is closely linked to their growth status [73].

Large diatoms like *Nitzschia palea* can sink rapidly, reaching speeds of up to 43 m/d or 500 µm/s [74]. These speeds might be variable, exhibiting an unsteady sinking pattern observed in large species like *Coscinodiscus wailesii* [75,76]. This behavior involves bursts of rapid sinking with a 1.5% change in mass density, coupled with periods of slow or near-zero speeds. Such fluctuating behavior among diatoms over timescales of seconds seems to confer a competitive advantage in nutrient assimilation [75]. Nevertheless, this fluctuating behavior appears absent under dark conditions, at least in the case of *Coscinodiscus wailesii* [77].

### 3.5. Lipids in Buoyancy Regulation

On the role of lipids in buoyancy regulation in phytoplankton, the literature reports seemingly conflicting results. Earlier studies generally considered lipids to play a minor role, despite their low intrinsic mass density, because buoyancy is largely governed by vacuoles - which can occupy a substantial fraction of cell volume (from ~30% in small diatoms to up to ~90% in larger ones) - and by dense structural components such as silica frustules [62,78,79]. More recent studies, however, indicate that the role of lipids is context-dependent. For example, Zhang et al. [80] reported a strong positive correlation between sinking speed (increasing nearly threefold) and lipid content in the nanoplanktonic diatom *Phaeodactylum tricornutum* under phosphate limitation or depletion, alongside substantial lipid remodeling, including replacement of phospholipids with sulfolipids [80]. In contrast, Wu et al. [81] showed that in the same species (*P. tricornutum*) under general nutrient starvation, lipid accumulation (particularly neutral lipid storage) contributes to a reduction in sinking speed. Taken together, these results suggest that lipids do not exert a uniform effect on buoyancy, but that their influence depends on the broader physiological and compositional changes accompanying different nutrient conditions [81].

### 3.6. Modified Stokes Formulation

Theoretical models have been developed to predict the maximum sinking rate of cylindrical and spherical diatoms with greater precision than conventional Stokes law [26]. Unlike the classical formulation, which assumes a homogeneous particle, these models are derived by modifying Stokes' law to account for the heterogeneous structure of diatoms, in which a dense siliceous frustule surrounds a less dense cytoplasmic interior. This leads to a size-dependent effective density, since frustule mass scales approximately with surface area while cytoplasmic mass scales with volume, thereby explaining the departure from the $r^2$ scaling predicted by Stokes' law. Based on this theoretical model [26] the sinking speed of spherical diatoms is given by:

$$v = \frac{2g}{9\eta}\left[\rho_{\mathrm{cyt}}\frac{(r-t)^3}{r} + \rho_{\mathrm{fr}}\frac{(3r^2t-3rt^2+t^3)}{r} - \rho_{\mathrm{w}}r^2\right] \quad (3),$$

where $\rho_{\mathrm{cyt}}$, $\rho_{\mathrm{fr}}$, and $\rho_{\mathrm{w}}$ denote the mass densities of the cytoplasm, frustule, and seawater, respectively, and $t = b_1 r^{a_1}$ represents the radius-dependent frustule thickness (with dimensions of length). Here, $a_1$ and $b_1$ are scaling coefficients, and each term in the brackets has the dimension of mass per unit length ($ML^{-1}$). The cell density is $\rho_{\mathrm{cell}} = \rho_{\mathrm{cyt}}\frac{V_{\mathrm{cyt}}}{V_{\mathrm{cell}}} + \rho_{\mathrm{fr}}\frac{V_{\mathrm{fr}}}{V_{\mathrm{cell}}}$, where $V_{\mathrm{cyt}}$, $V_{\mathrm{fr}}$, and $V_{\mathrm{cell}}$ represent the volumes of the cytoplasm, the frustule, and the cell. One suggestion for future research is to incorporate the significant influence of vacuole mass density ($\rho_{\mathrm{vac}}$ = 1017.6 kg.m$^{-3}$ [68]), due to their considerable volume and role in positive buoyancy [25,79,82], into Equation (3). This factor reduces overall cell density, thereby facilitating positive buoyancy [83]. The approach to do this is to add a new term in the form of $\rho_{\mathrm{vac}}\frac{V_{\mathrm{vac}}}{V_{\mathrm{cell}}}$ into Equation (3). To determine the volume of the vacuole relative to the cell volume, data provided in [79] for diatoms can be used. By fitting a function to these data, the ratio $\frac{V_{\mathrm{vac}}}{V_{\mathrm{tot}}}$ for different cell volumes can be obtained. For example, in a cell with a volume of $5.4\times10^4$ µm$^3$, approximately 90% of the protoplast volume is occupied by the vacuole [79], making the vacuole-to-cell volume ratio significantly higher than the relative contributions of the frustule and the remaining cytoplasm.

How much variation in density is possible, and how long does it take for the buoyant cell to achieve this? According to [84], *Pyrocystis noctiluca,* a non-flagellated dinoflagellate, experiences a six-fold increase in cellular volume within a span of less than 10 minutes. This process involves the calcium-activated influx of *fresh* water through aquaporin channels, followed by transport to the vacuole, resulting in the reversal of gravitational sedimentation [84]. This creates a situation where the cell mass density, $\rho_{\mathrm{cell}} = \rho_{\mathrm{cyt}}\frac{V_{\mathrm{cyt}}}{V_{\mathrm{cell}}} + \rho_{\mathrm{vac}}\frac{V_{\mathrm{vac}}}{V_{\mathrm{cell}}}$, becomes less than the mass density of seawater.

### *3.7. Buoyancy Regulation Efficiency*

Efficiency of buoyancy regulation machinery has largely been overlooked in the literature and will be addressed in Section 5 of this review. We express this efficiency through the energy dissipated in cellular mass density regulation, including processes involving vacuoles, gas vesicles, and frustules. Among the few aspects of buoyancy regulation efficiency that have been investigated in the literature is the effect of shape factor. Although shape factor is not included in Equations 2 and 3, its influence has been examined elsewhere. Non-spherical geometric shapes generally sink at a slower rate compared to spheres of the same volume [18,85]. However, tear-drop-shaped phytoplankton (e.g. *Rhodomonas* and *Gymnodinium*) serve as an exception, sinking faster than their spherical counterparts [18,85]. To the best of our knowledge, there is no clear physical or physiological principle that explains the functional efficiency of different shapes or how shape influences the energetic cost of buoyancy regulation.

## 4. Phytoplankton Migration: Survival Strategies

Why is the cost of motility a time-dependent, dynamic process? Because phytoplankton engage in migratory behaviors that expose them to constantly changing environmental conditions, which directly and indirectly influence their motility costs and benefits. These varying conditions include changes in temperature, light, nutrient availability, viscosity, predation and viral risks, and turbulence. Therefore, before examining the costs of migration, it is essential to understand the fundamentals of phytoplankton migration. In this section, we discuss the time-scale and length-scale of vertical migration for both relatively slow and fast phytoplankton, and their role in net primary production. We examine the velocity at which migrating phytoplankton reach their target depth, the impact of increasing speed on the accumulation of species at specific depths, and the influence of turbulence on diversity. Additionally, we explore the strategies phytoplankton employ in turbulent waters, such as chain formation, division into sub-populations, and changes in orientation, direction, and shape.

Phytoplankton primarily migrate by orienting themselves through chemotaxis, phototaxis, and gravitaxis - movements in response to chemical gradients, light, and gravity, respectively. When species migrate either towards the ocean's surface, where light intensity is higher, or to deeper depths with greater nutrient concentrations, there exists the potential for an increase in their rates of photosynthesis or nutrient uptake [86,87]. The light-rich and nutrient-rich regions are typically separated by distances of 30-120 m, and relatively fast phytoplankton species are capable of vertical diel migration within these regions [86,87]. Organisms that migrate diurnally to the surface during the light period may form a subsurface layer when faced with nutrient limitations (e.g., nitrate), where light and nutrient availability are optimally balanced [88]. For a comprehensive review of the mechanisms underlying phytoplankton layer formation - including straining, convergent swimming, buoyancy regulation, gyrotactic trapping, *in situ* growth, and intrusions - see [89].

Phytoplankton species use different depth-regulation strategies - mixing, migration, and layer formation - based on their motility and adaptation to specific hydrographic regimes [90]. These strategies involve ecological trade-offs in physiology and behavior, including photoacclimation, motility control, nutrient storage and uptake, and photoprotection [90]. Together, these traits define species-specific ecological niches and help explain the success of certain phytoplankton in forming harmful algal blooms under varying oceanographic conditions.

During vertical migration, the species may also exhibit limited horizontal movement, relative to vertical migration, through diffusion alone, typically spanning up to 10 centimeters, as observed in ideal lab experiments in the absence of water currents [91]. This horizontal displacement is relatively insignificant compared to their predominant vertical movement; however, tidal currents can significantly influence the horizontal movement of species [92].

Contrary to the common assumption that most phytoplankton species lack migratory behaviors due to their relatively slow speed, Wirtz and Smith [63,93] reported that the observed depth-dependent profile of chlorophyll-a concentration [94] in the oligotrophic ocean is attributed to long-term (days to week) slow vertical migration via sinking/rising or swimming processes. This slow form of active mobility, which is considerably more challenging to track compared to diel migration, contributes to 7-28% of the ocean's net primary production [63,93]. The variation in the nutrient-to-carbon ratio at identical ocean depths is suggested to result from distinct migration histories, involving movement either from the upper surface ocean to deeper depths or vice versa [93].

Klausmeier and Litchman [95] developed a dynamic mass-balance model that considers swimming/sinking speed as a parameter, integrating factors such as growth rate under light and nutrient limitation, loss rate, eddy diffusion, and active movement to predict the vertical distribution of phytoplankton in poorly mixed water columns. Active movement is expressed as a term in a time-dependent biomass-balance equation, represented as a one-dimensional gradient of the product of speed ($v$), biomass ($B$), and the gradient of the growth rate ($\mu$), written as $\frac{\partial}{\partial z}\left(v\frac{\partial \mu}{\partial z}B\right)$. Their findings demonstrate that as phytoplankton speed increases, they tend to concentrate within a narrow layer at a specific depth, where they experience an equal limitation of light and nutrients. This depth represents an evolutionarily stable strategy [95].

Turbulence is an inherent characteristic of oceanic waters that significantly affects phytoplankton, impacting their mortality rate, motility costs, distribution, and diversity [96]. As a result, phytoplankton must develop strategies to adapt to turbulent conditions. The fundamental physics governing some idealized strategies of a model swimmer, such as swimming towards safer points, minimizing swimming time and path in parallel currents, source currents, and sink currents, is explored in [97]. However, migrating phytoplankton perform more complex tasks, such as reorienting their swimming trajectory in response to turbulence [98] and engaging in collective behavior by forming elongated chains, as observed in diatoms, cyanobacteria, and dinoflagellates within the water column [98–100].

The formation of chains enhances swimming/sinking speeds particularly in weak to moderately turbulent water flows [100]. This occurs because as the number of cells in a chain increases, the collective propulsive force outweighs the hydrodynamic drag acting on the chain [101]. The swimming speed of the chain can be determined by considering both the number of cells comprising the chain and the chain's shape factor [100] as following:

$$v_c^{(n)} = n^{\left(\frac{2}{3}\right)} \frac{v_c^{(1)}}{K} \quad (4),$$

where $v_c^{(n)}$, $n$, $v_c^{(1)}$, and $K$ represent, respectively, the chain swimming speed, the count of rigidly attached spherical cells, the swimming speed of a single species cell, and the shape correction factor. While Equation (4) describes how swimming speed scales with chain length and geometry, estimating the metabolic cost of swimming requires applying Stokes' law in conjunction with an assessment of the chain's mechanical efficiency - a key factor influencing total energy expenditure that, to the best of our knowledge, remains unexplored.

In turbulent hydrodynamic conditions, it has been reported that dinoflagellates and raphidophytes may divide into two subpopulations within approximately 30 minutes [102]. One group propels upwards while the other navigates downwards, facilitated by geometric shape adjustments to their motility mechanisms [102] which directly affect motility efficiency. To model the dynamic morphological changes, particularly the fore-aft asymmetry discussed in [102], it is necessary to investigate the center of buoyancy, center of mass, center of hydrodynamic stress, and the organism's orientational stability. While such rapid responses to severe environmental shifts may expose the microorganism to elevated cellular stress (characterized by nitric oxide, NO, production) and incur significant costs, they underscore the species' advanced survival strategies.

## 5. Trade-offs of Phytoplankton Movement: A Cost-Benefit Perspective

Swimming and buoyancy regulation have metabolic costs for both the construction (e.g., biosynthesis of flagella) and operation. We can analyze our understanding of the factors influencing migratory behavior by analyzing these costs and comparing them to the benefits of motility.

### *5.1. Benefit Analysis*

Motility enables phytoplankton to reorient along the light vector through phototaxis [103], enhancing photosynthetic efficiency, and to navigate toward higher nutrient concentrations via chemotaxis [104,105]. Both processes rely on complex signal transduction pathways [49]. In most ecological settings, these taxis-driven processes constitute the dominant advantage of motility, whereas hydrodynamic enhancement of nutrient uptake plays a comparatively secondary role. In addition to these taxis-driven benefits, the motility apparatus (such as specific flagellar arrangements) [31] can generate feeding currents that improve nutrient access. Jumping-like movements also contribute significantly to overcoming diffusion limitations and enhancing nutrient uptake [46]. Unlike motility costs, which can often be expressed as instantaneous energetic expenditures, the benefits of motility are inherently context-dependent and emerge over longer temporal and spatial scales. These benefits depend strongly on environmental structure, patch persistence, turbulence, and loss processes, and therefore do not admit a universal quantitative metric analogous to cost estimates. The Sherwood number provides a rare case where the benefit of motility can be quantified in a relatively general and organism-independent manner, which is why it is emphasized here.

The Sherwood number (Sh) is a dimensionless number that quantifies how much faster phytoplankton can take up nutrients by moving compared to relying solely on passive diffusion. This enhancement of nutrient uptake during swimming or sinking/rising results from advection and is expressed as [106,107]:

$$\mathrm{Sh} = \frac{Q}{4\pi r D C_\infty} \quad (5),$$

in which $Q, r, D$, and $C_\infty$ represent, respectively, the nutrient flux, cell radius, diffusion coefficient, and nutrient concentration at a distance far from the cell. A stationary cell is characterized by an Sh of 1 while an Sh of 1.2 signifies a 20% enhancement in nutrient uptake rate attributed to swimming/sinking. For example, empirical investigations of the flow field around the dinoflagellate *Dinophysis acuta* (79 µm in length, 54 µm in width, 104 µm/s swimming speed, $D = 10^{-9}$ m²/s, with constant nutrient concentration far from the cell and zero nutrient concentration at the cell surface) show that swimming contributes to a substantial 75% increase (Sh = 1.75) in nutrient uptake rate compared to diffusion alone [107]. Jumping behavior in photosynthetic *Mesodinium rubrum* significantly enhances nutrient uptake by increasing the Sherwood number, but only for a very brief period [46]. A one-beat short jump in Antarctic strains can raise Sh to 2.5, while multiple-beat long jumps reach up to Sh = 3.3. In temperate strains, long jumps peak at Sh = 3.9, indicating more efficient nutrient transport. Additionally, brief episodes of fast sinking are reported to allow *Coscinodiscus wailesii* diatoms to enhance nutrient uptake by 170% compared to when they sink at their usual rate [75].

To calculate the Sherwood number, one can use established correlations between the Sherwood and Péclet numbers, where $\mathrm{Pe} = \frac{rv}{D}$ is a dimensionless measure of the relative importance of advective to diffusive transport. Such correlations have been developed for spherical steady swimmers [108], monoflagellated organisms [106], and biflagellated organisms [20]. An example of such an empirical relationship for a sinking sphere found in the literature is [109]:

$$\mathrm{Sh} = \frac{1}{2}\left(1 + (1 + 2\mathrm{Pe})^{1/3}\right). \quad (6).$$

Given that the speed and characteristic radius of phytoplankton species are documented in the literature, it is straightforward to estimate the Sherwood number and consequently evaluate the enhancement in nutrient uptake attributed to swimming or sinking using Equation (6). For example, for the spherical-shaped sinking coccolithophore *Calcidiscus leptoporus* with a radius of approximately 10 µm and a mean sinking rate of 4.3 m/d [74], the Peclet number is calculated as: $\mathrm{Pe} = \frac{10\times10^{-6}[\mathrm{m}]\times4.3\times(86400)^{-1}\left[\frac{\mathrm{m}}{\mathrm{s}}\right]}{10^{-9}\left[\frac{\mathrm{m}^2}{\mathrm{s}}\right]} \cong 0.5$. This gives the Sherwood number as $\mathrm{Sh} = \frac{1}{2}\left(1 + (1 + 2\times0.5)^{1/3}\right) \cong 1.13$ indicating a 13% increase in the nutrient uptake rate.

As an approximation, it can be shown that under nutrient-limited conditions, the ratio of the Sherwood number to the dimensionless relative flagellar cost $C_{\mathrm{rel.}}$ (which includes both construction and operating costs) yields the relative growth rate $\mu_{\text{rel.}}$ - that is, the growth rate of a motile cell relative to a stationary one [24]:

$$\mu_{\text{rel.}} = \frac{\text{Sh}}{1+C_{\text{rel.}}}. \qquad (7).$$

According to the estimate outlined in Equation (7), eukaryotic flagellated microswimmers with a volume exceeding 1000 μm³ encounter an increase in their growth rate, while those with a volume less than 100 μm³ find nutrient gathering by flagella to be uneconomical, since its cost outweighs its benefits [24]. This suggests that there might be other benefits to employing flagella in relatively small species, besides enhancing immediate nutrient uptake rate.

### *5.2. Cost Analysis*

Motility costs encompass all costs associated with the synthesis, remodeling, degradation, and operation of motility machinery, spanning the period from the initiation of movement to its cessation, with magnitudes modulated by environmental conditions. Because protein degradation incurs a substantially lower ATP cost than protein synthesis - generally a few percent and at most ~10% in high-turnover systems [110,111] - we neglect degradation costs in this review. Thus, motility costs include: 1) the cost of constructing motility-related machinery, such as visible structures like flagella or non-visible components found in cyanobacteria, diatoms, etc.; 2) the minimum power required for swimming (determined by the drag force exerted on the organism due to viscosity and pressure) and for buoyancy regulation (based on molecular mechanisms of shape and mass density adjustments); 3) the cost related to the efficiency of motility-related internal and external machinery; and 4) the costs associated with any environmental changes during migration. Some costs occur at specific points during the cell cycle - such as the synthesis and remodeling of motility machinery - and are considered one-time construction expenses, typically measured in joules or ATP per cell. In contrast, other costs are incurred continuously over time (operational costs) and are measured in joules or ATP per second per cell. The energy released from hydrolyzing one mole of ATP ranges from 28 kJ/mol under standard conditions to approximately 55 kJ/mol in *E. coli* during aerobic exponential growth, and can reach up to 72 kJ/mol in *Homo sapiens* muscle during intense exercise. This variation depends primarily on intracellular pH and magnesium ion concentration, as $Mg^{2+}$ stabilizes ATP by binding to its phosphate groups [112]. For all calculations in this study, we adopt a representative value of 50 kJ/mol.

#### *5.2.1. Construction Cost: Flagellated Phytoplankton*

One notable expense associated with the phytoplankton swimming process is the construction cost of the flagellum. While it might be assumed that this is a one-time occurrence and therefore not a substantial energy fraction over the lifespan, research indicates that microorganisms can lose their flagella either accidentally or through programmed mechanisms. For instance, bacteria have the ability to remove or deactivate their flagellum in conditions of nutrient starvation [113,114]. Consequently, it is important to account for a flagellum loss rate when studying phytoplankton swimmers.

The construction cost refers to the expenses associated with building the proteins and lipids necessary for the formation of a flagellum. This is calculated by considering the number of proteins, the length of the proteins in terms of amino acids, and the average energy cost per amino acid as proposed by [24]. Additionally, because eukaryotic flagella are enclosed by a membrane, the construction cost of the membrane should also be included. According to the estimation provided by Schavemaker and Lynch, without accounting for flagella loss rate, the costs of constructing flagella relative to the cost of constructing the whole cell in eukaryotes, assuming a spheroidal cell shape, are suggested to range between 0.037% and 44%, with a median value of 3% [24]. The absolute cost of constructing flagella rises with cell volume because the total flagella length (product of the number and length of flagella) increases with cell volume, albeit with varying exponents among different groups. However, relative to the total construction cost of the cell, the cost of flagella construction tends to decrease as cell volume increases [24].

The cost associated with any modification to the flagellum, such as adding a single amino acid to the protein's structure, varies depending on the type of protein involved. This range extends from the relatively low cost of inner-arm dyneins to the higher cost of tubulins, in relation to the overall cell construction cost [24]. The cost assessment of modifications in the flagella is important because the length of flagella varies within cells due to biological fluctuations connected with LF genes (specifically LF1 [115,116]), which regulate the flagellum length-control machinery and lead to the addition or removal of a large number of tubulins [115]. The system regulating flagellar length in *Chlamydomonas reinhardtii* displays fluctuations significantly greater than measurement error, with a variability coefficient ranging from 10% to 20% [115]. Therefore, the cost associated with these within-cell flagella length fluctuations [115], which involve the removal and addition of amino acids from/to proteins constructing the flagella, is important, as estimated in [24]. However, there should always be a break-even point where further increases or decreases in flagella length become ineffective for the cell. The length should be adjusted to maintain a balance between costs and benefits, ultimately leading the cell to adopt an optimal intermediate length.

A microswimmer with a higher construction cost of flagella (due to having more and/or larger flagella) possesses the capability to achieve greater swimming speeds, as the log-log experimental data shows (albeit with a low $R^2 \sim 0.35$) [24], leading to increased drag force and power consumption in a viscous environment.

### *5.2.2. Construction Cost: Swimming Cyanobacteria*

Cyanobacteria such as *Synechococcus* and *Synechocystis* sp. swim using non-visible machinery including protein motors that generate traveling waves on the cell surface, but little is known about the construction cost of the components involved in the propulsion of these organisms. Here, we provide a biological overview of the construction cost of these non-visible motility apparatuses. However, precise estimation of the associated costs remains a subject for future studies.

The essential components for the swimming of *Synechococcus* include the crystalline S-layer (surface layer) [117] composed of elongated subunits tilted approximately 60° with respect to the cell wall, SwmA glycoproteins [118], SwmB polypeptides [119], and protein motors associated with the peptidoglycan layer [33]. To estimate the construction cost of these motility-related components, one needs to determine the number and length of the subunits of the S-layer, SwmA, SwmB, and motor proteins. To the best of our knowledge, the copy number of all these components is not available in the literature.

The construction cost of motility-coupled components in *Synechocystis* sp. is linked to the copy number of type IV pilus (T4P) machines. These machines are composed of major and minor pilins, secretin, alignment proteins, platform proteins, assembly ATPase, and retraction ATPase [40]. To estimate the construction cost of the T4P in *Synechocystis* sp., we refer to the PDB ID 3JC8, which represents the "Architectural model of the type IVa pilus machine in a piliated state" of *Myxococcus xanthus* [120]. This structure, determined using electron microscopy, consists of 37,468 amino acid residues [120]. Recent measurements indicate that the average copy number of T4P machines in *Synechocystis* sp. varies between 15 and 20 per cell [121]. The final piece needed to calculate the total construction cost is the ATP hydrolysis cost per amino acid, which is provided below.

Constructing a single amino acid under aerobic conditions with glucose as the primary carbon source involves both a direct ATP cost and an opportunity cost. The opportunity cost refers to the energy the cell forgoes by diverting a metabolic intermediate into biosynthesis rather than fully oxidizing it for energy. For example, synthesizing alanine from pyruvate requires a direct expenditure of 2.5 ATP equivalents. However, if the cell had instead continued metabolizing pyruvate through energy-generating pathways, it could have yielded an additional 12.5 ATP equivalents. Therefore, the total cost of producing alanine is the sum of these two values: 15 ATP equivalents. Detailed breakdowns of the direct, opportunity, and total ATP costs for all 20 amino acids are provided in Foundations 17.2 of [49]. We downloaded the FASTA sequence file for PDB ID 3JC8 to count the total number of each amino acid type. By multiplying these counts by the ATP hydrolysis cost per amino acid type (as provided on page 435 of [49]), we obtain an average (total) construction cost of 26.3 ATP hydrolysis events per amino acid for this specific T4P machine. Using this value, the total construction cost of the T4P machines can be estimated as: 20 [machines] × 37,468 [residues per machine] × 26.3 [ATPs per residue] = 19,708,168 ATPs, or approximately $2.0 \times 10^7$ ATP hydrolysis events.

Given the potential differences in the number of residues between T4P of *Myxococcus xanthus* and *Synechocystis* sp., we also propose an alternative approach. We use the reported size of T4P machines in *Synechocystis* sp. [121] to estimate their volume. Using the average volume per amino acid residue ($1.33 \times 10^{-10}$ µm³ [24]), we can then determine the number of residues per T4P machine. The average length and width of T4P machines in *Synechocystis* sp. are reported to be 2.5 µm and 0.007 µm, respectively [121]. Assuming the T4P structure is cylindrical, this gives an approximate volume of $9.6 \times 10^{-5}$ µm³. Dividing this volume by the average volume per amino acid residue (assuming a packing fraction of 100%, i.e., that amino acids completely fill the volume of the cylindrical T4P structure) yields approximately 723,393 residues per T4P machine in *Synechocystis* sp., which is about 20 times higher than the residue count derived from the PDB ID of the *M. xanthus* T4P. Therefore, the total construction cost of the T4P machines in *Synechocystis* sp. can be calculated as: 20 [machines] × $7.2\times10^5$ [residues per machine] × 26.3 [ATPs per residue] = $3.8\times10^8$ ATP hydrolysis events. Assuming that T4P machines are the sole contributors to motility in *Synechocystis* sp., our estimated construction cost agrees with the median construction cost reported for bacterial flagella in the literature [24] and accounts for a small fraction of the overall cellular construction cost.

### *5.2.3. Construction Cost: Buoyancy Regulation Machinery*

The construction cost associated with buoyancy regulation in phytoplankton varies with the number and size of proteins associated with the synthesis of gas vesicles in cyanobacteria [12], the need to syntheses organic osmolytes [68], the construction cost of pores and channels that mediate the active influx of water from the cytosol into the vacuole [68], and the replacement of heavy ions with light ions in the vacuole [27,68]. In diatoms, the energy investment in frustule formation, which is strongly related to sinking, entails silica precipitation, cytoskeletal organization of silica-rich vesicles, silica metabolism regulated by genes, and assembly costs [27,28]. The calcification and formation of the coccolith layer in coccolithophores play a similar role in contributing to the sinking patterns of these microorganisms [122]. One challenge in determining the construction cost of the components regulating phytoplankton buoyancy is that these components serve multiple purposes, making it difficult to estimate the fraction of energy expenditure specifically related to buoyancy regulation.

### *5.2.4. Construction Cost: Buoyancy Regulating Cyanobacteria*

For gas vesicles to provide buoyancy, 3 to 10 percent of the cell volume should be occupied by gas vesicles, depending on their mass density [12]. Approximately 10% of this volume is composed of the protein wall, which has a mass density of around 1300 kg/m$^3$, while the remaining volume is gas space, with a mass density of about 1 kg/m$^3$ [12]. If a cell of *Escherichia coli* size and density were hypothetically to achieve neutral buoyancy in freshwater via gas vesicles (used here purely as a reference scale, since *E. coli* does not naturally produce gas vesicles), what would be the associated construction cost? Walsby [12] estimated that the one-time energetic investment required to synthesize sufficient gas-vesicle protein for neutral buoyancy in an *E. coli*–sized cell ($V$ = 0.896 μm$^3$) is approximately $2.84\times10^{-10}$ J, corresponding to $3.42\times10^{9}$ ATP equivalents. According to Raven and Lavoie [25], the energy cost of synthesizing gas vesicles in cyanobacteria is $1.3 \times 10^{-3}$ of the total cellular synthesis cost. As described earlier in Section 3, buoyancy regulation in cyanobacteria such as *Microcystis aeruginosa* is governed not only by gas vesicles but also by diel metabolic processes, whereby carbohydrates are synthesized under illumination via photosynthesis and subsequently consumed in the dark through respiration [64]. To the best of our knowledge, the energetic cost of carbohydrate accumulation as ballast for buoyancy regulation in cyanobacteria has not been quantified in the literature; however, the costs associated with other ballast materials in diatoms and coccolithophores are discussed below.

### *5.2.5. Construction Cost: Diatoms and Coccolithophores*

Lavoie *et al.* [67] reported that buoyancy-modulating mechanisms in aflagellated phytoplankton (including coccolith formation, silica cell wall formation, polysaccharide storage, and DMSP synthesis, with DMSP influencing buoyancy via replacement of quaternary ammonium osmolytes and slight changes in cell density) span a wide energetic range, from very low (~0.1%) to substantial (~12%) fractions of the total absorbed light used for growth under low-nitrogen conditions. In *E. huxleyi*, coccolith formation and polysaccharide storage impose non-trivial photon costs (≈2.4–9.5% of the cellular photon budget), whereas DMSP synthesis is comparatively inexpensive in photon terms (≈0.43–0.47%). In *T. pseudonana*, silica cell wall formation and polysaccharide storage incur similar photon costs (≈3.1–12%), but silica biomineralization can additionally require a substantial fraction of cellular nitrogen (≈24–40% of total N). In the diatom *E. rex*, polysaccharide storage and DMSP synthesis account for ≈4.3% and ≈2.4–3.5% of the total photon cost, respectively, while silica cell wall formation may be potentially costly in nitrogen.

### *5.2.6. Operational Costs: Flagellated Phytoplankton*

As discussed in Section 2, Stokes' law (Eq. 1) provides the minimum power required for propulsion of an ideal spherical microswimmer, thereby establishing a lower bound on the operational cost of swimming. In practice, however, the dominant contribution to this cost is governed by the efficiency of the swimming machinery, which determines how much additional energy must be expended beyond this theoretical minimum. Mechanical efficiency refers to how effectively a swimmer converts its energy into motion, quantified as the ratio of the minimum power required to pull the organism - based on its drag at the average swimming speed over one beating cycle - to the average amount of power it actually delivers to the fluid during that cycle [47,50]. The swimming efficiency of flagellated unicellular organisms, whether eukaryotes or bacteria, is generally below 3% [13,19–21,47,123–125]. For example, a low efficiency of 1% means that the actual power required for swimming is 100 times greater than the minimum estimated using Stokes' law (Equation 1). This energetic cost may be substantial relative to the organism's total metabolic rate, especially under nutrient-limited conditions.

Swimming efficiency can be divided into external and internal efficiency. External efficiency encompasses energy dissipation related to the swimmer's external characteristics, such as body geometric shape [17], flagellum-to-body length ratio [20,21], and flagellum slenderness [20,21], all of which are affected by the hydrodynamic viscosity of the environment. In contrast, internal efficiency pertains to energy dissipation within the organism, such as the elastic bending of the axoneme [22,23]. The external efficiency associated with the physical properties of flagella has been reported to be 0.8% for the green alga *Chlamydomonas* [20] and 1.4% for *Tetraflagellochloris mauritanica* [21].

As described above, the external efficiency depends, in part, on the shape of the microswimmer. Ignoring the trade-off between microswimmer shape, navigation cost, and optimal navigation time [126], higher external efficiency values are observed for prolate-shaped microswimmers. These values decrease as the shape transitions towards more compact forms (such as spherical or cubic), with the lowest values found in oblate shapes [17,127,128]. This assessment is based solely on cell shape and excludes motility machinery such as flagella; however, differences in flagellar placement across cell shapes undoubtedly complicate the problem. The maximum external efficiency ($\xi_m$), corresponding to the minimum external dissipation, for various geometric shapes is given by the following expression. In Equation (8), $R_{\mathrm{PS}}$ represents the drag coefficient of a perfect-slip rigid body, for which the boundary is impermeable and the tangential shear stress vanishes, whereas $R_{\mathrm{NS}}$ denotes the drag coefficient of a no-slip rigid body, for which the fluid velocity at the boundary matches that of the boundary itself [17]:

$$\xi_m \leq 1 - \frac{R_{\mathrm{PS}}}{R_{\mathrm{NS}}} \quad (8).$$

The drag coefficient for spheroidal bodies under both no-slip and perfect-slip boundary conditions has been derived explicitly in the literature [17,129]. These formulations allow the maximum efficiency to be computed as a function of the spheroid's aspect ratio, as illustrated in Figure 3 of [17]. To simplify the estimation of shape-related efficiency, a sigmoid function can be fitted to Figure 3 of [17], allowing efficiency to be estimated using only the shape's aspect ratio. This approach is applied as a working example in Section 5.3.

The theoretical model of Ref. [130] shows that, for microswimmers using tangential surface actuation - where the surface slides fluid backward, like surface treadmilling, to move forward - the power required for swimming can vary by more than six orders of magnitude depending on the shape's aspect ratio. Based on this theory [130], and considering empirical data on the elongation and aspect ratio of phytoplankton species [131], one finds that the external energy dissipation during swimming in certain *Ceratium* species (a genus of dinoflagellates characterized by an aspect ratio of 0.025 and oblate elongation) is exceptionally higher than in certain *Closterium* species, a genus of charophyte green algae characterized by an aspect ratio of 112.83 and prolate elongation, with the difference reaching roughly six orders of magnitude.

The examination of over 5700 phytoplankton species' shapes spanning 402 genera reveals they can be categorized into 38 distinct geometric forms [131]. This investigation indicates that intermediate-sized species exhibit the most significant shape variation, while small and large cell sizes tend to be predominantly compact (spherical or cubic). Among the various shapes investigated, the authors reported that 46% are considered prolate, 38% compact, and 16% oblate [131]. Prolate shapes are elongated cells with an aspect ratio greater than 1, moving primarily along their longest axis (like a cigar). Oblate shapes are flattened cells with an aspect ratio less than 1, which can move edge-on, with their flat faces aligned with streamlines of the surrounding flow (lentil-like) [132]. Compact shapes have an aspect ratio close to 1 and show no strong preference in movement direction. A complicated yet valuable endeavor involves applying the theoretical framework pioneered by [127] to calculate the external power dissipation during swimming for these 38 shapes [131].

Swimming internal efficiency refers to the effectiveness of the internal propulsive layer, axoneme, of the phytoplankton swimmer. The primary sources of energy dissipation in the internal propulsive layer are flagellum bending, sliding of microtubules, and internal viscosity [22,23]. Experiments on the ATP consumption rate per beat cycle of the flagellum in *Chlamydomonas reinhardtii* reveal that the energy dissipation due to the elastic bending of the axoneme is an order of magnitude greater than the internal hydrodynamic energy dissipation [22]. Although Chen *et al.* [22] did not report swimming efficiencies for *Chlamydomonas*, they defined hydrodynamic efficiency ($\varepsilon_{\mathrm{hydro}}$) and chemomechanical efficiency ($\varepsilon_{\mathrm{chemo}}$), and quantified the overall swimming efficiency ($\varepsilon_{\mathrm{swim}}$) for sea urchin sperm. Given the conserved dynein-driven 9+2 axonemal structure across eukaryotic flagella, including those of phytoplankton, these estimates provide a relevant reference for motility energetics. They found that $\varepsilon_{\mathrm{swim}}$ ranges from $\varepsilon_{\mathrm{swim}} = 0.34 \times 0.004 = 0.001$ to $\varepsilon_{\mathrm{swim}} = 0.6 \times 0.013 = 0.008$, depending on the viscosity.

#### *5.2.7. Operational Costs: Swimming Cyanobacteria*

The detailed swimming internal efficiency of *Synechococcus* and *Synechocystis* cyanobacteria remains less explored. However, it is evident that any energy dissipation linked with motility mechanisms should be taken into account. For instance, in *Synechococcus*, components such as the S-layer and protein motors moving along helical paths contribute to energy dissipation, while in *Synechocystis*, the type IV pilus could play a similar role. In *Synechococcus* cyanobacteria, the subunits of the S-layer form a two-dimensional lattice with varying spacing, typically ranging from 5 to 22 nm, resulting in oblique, triangular, square, or hexagonal shapes [133]. These structural variations, unique to each species, may lead to different ranges of energy dissipation. This dissipation likely occurs as motor proteins make frictional contact with the S-layer and move beneath it along a helical path, generating traveling waves on the organism's surface [33]. Since these motor proteins are driven by proton motive force [33], another source of dissipation likely occurs in this process, presumably in the form of proton leakage, as extensively reported in the membrane of other organelles like mitochondria [134–136]. In type IV pilus machines in *Synechocystis*, one reported source of inefficiency is related to the retraction process, mostly in the absence of ATPase [40,137].

One might question the significance of swimming costs in cyanobacteria, given their typically small size and slow swimming speeds - although some species, such as those in the order *Nostocales*, can reach millimeter scales. According to Stokes' law, this would suggest that their swimming incurs minimal energetic cost. However, two important points should be considered: (1) the swimming efficiency of cyanobacteria remains unknown, and (2) the energetic cost of swimming is a relative measure - it must be evaluated in relation to the organism's metabolic rate. Since smaller organisms tend to have lower metabolic rates, even small swimming costs may represent a substantial fraction of their energy budget. The same principle applies to the cost of buoyancy regulation in cyanobacteria, which is explored in the following section.

#### *5.2.8. Operational Costs: Escape-Jumping Phytoplankton*

As briefly discussed in Section 2, some phytoplankton swimmers exhibit jumping-like behavior either to escape predators or to enhance nutrient uptake. However, these jumps come at a high energetic cost. First, according to Stokes' law, if an organism increases its speed fivefold during a jump, the power required for swimming rises by at least 25 times. Second, additional energy is lost through dissipative processes during each jump. Using computational fluid dynamics (CFD), Jiang [138] demonstrated that although a typical jump lasts only 18.6 milliseconds, it can consume between 8% and 185% of the respiration rate of the photosynthetic *Mesodinium rubrum* in that small time interval, depending on the ciliary arrangement and forcing scheme. Nevertheless, the author concluded that if only one jump occurs per second, the average energetic cost of jumping over one second would be relatively small compared to the respiration rate [138].

### *5.2.9. Operational Costs: Buoyancy Regulating Phytoplankton*

Although the operational costs of buoyancy regulation (i.e., the underlying molecular mechanisms) have been extensively studied in the literature, their efficiency and the associated dissipative processes have largely been overlooked. The shape-dependent external efficiency of sinking species appears to be more complex than that of swimmers [74]. Non-spherical sinkers (both single cells and colonies) generally exhibit a slower sinking rate compared to their equivalent spherical counterparts, except for tear-drop-shaped sinkers [18,139]. To the best of our knowledge, there is no solid physical theory in the literature explaining why the shape-related external efficiency differs between swimmers and sinkers.

The correlation among shape, speed, ocean depth, nutrient concentration, and light availability is significant. For instance, deep waters often exhibit low-light and high-nutrient conditions concurrently. There is an augmentation in cell elongation with increasing depth, regardless of whether the waters are poorly mixed or well-mixed [131]. This is presumably because phytoplankton with elongated shapes are adept at arranging chloroplasts along their cell surfaces, thereby enhancing light absorption, particularly at greater depths [140]. In marine systems, seasonal variations in community composition lead to a predominance of oblate-shaped diatom species in spring and autumn, whereas prolate-shaped species dominate during summer [141].

The internal efficiency of sinking and rising in phytoplankton remains largely unknown. This is likely because these species have traditionally been considered passive motile microorganisms in physical models. We propose classifying their motility as active, albeit non-visible. Using this new classification, it is essential to define, calculate, and formulate the internal efficiency of phytoplankton's sinking and rising movements. This internal efficiency can be defined as any dissipation occurring in contact with the components that facilitate sinking/rising. The buoyancy regulation strategy can potentially be more energy-intensive than swimming, owing to the diverse modulatory processes a cell must undertake to move both downward and upward [25,67].

### *5.2.10. Operational Costs: Buoyancy Regulating Cyanobacteria*

Gas vesicles, which mediate buoyancy regulation in prokaryotes such as cyanobacteria and haloarchaea, consist of a gas-filled space enclosed by a protein wall primarily composed of proteins like GvpA and GvpC [12,142,143]. The efficiency of gas vesicles in providing buoyancy is proposed to be influenced by the interaction of their geometrical properties, elastic modulus, and turgor or hydrostatic pressure [144]. The geometrical shape of gas vesicles changes over time, starting as small biconical structures at formation and developing into cylindrical and spindle shapes at maturity [143].

This efficiency of gas vesicles is qualitatively defined by the volumetric ratio of the gas space, $V_i$, to the surrounding proteins, $V_w$, which indicates how much buoyant force is generated relative to the structural material [144]. As the radius and length of the cylindrical-shaped vesicle increase, the efficiency improves because a larger gas volume provides more buoyancy with relatively less protein weight. However, this improvement in efficiency eventually saturates with increasing length [12,144]. Under normal conditions, gas vesicles can achieve efficiencies as high as 90% of the theoretical maximum, defined as $\frac{\left(\frac{V_i}{V_w}\right)_{\text{measured}}}{\left(\frac{V_i}{V_w}\right)_{\text{theor.max.}}} \times 100$, formulated for infinitely long gas vesicles [144]. However, changes in hydrostatic pressure, such as those experienced in deep ocean environments or under high light intensity, can reduce this efficiency and may cause gas vesicles to collapse, disrupting buoyancy regulation in cyanobacteria.

For freshwater cyanobacteria, although most lakes are relatively shallow (having a global mean depth of ~15 m, with reported ranges from ~0.5 m to ~700 m [145]) and thus do not generally impose the high hydrostatic pressures characteristic of marine environments, Abeynayaka et al. [146] demonstrated experimentally that exposure of *Pseudanabaena galeata* to elevated pressures (mimicking hypolimnetic conditions of lakes) results in reduced intracellular ballast and complete loss of gas vesicles, thereby preventing cells from maintaining buoyancy over the duration of the experiment [146]. If the cell survives, reforming gas vesicles is a resource-intensive process. The ability of gas vesicles to withstand pressure depends on the elastic modulus of their structure, necessitating a mechanical analysis of stress and strain [147].

### *5.2.11. Operational Costs: Coccolithophores*

Calcification associated with the calcified shells (coccospheres) of coccolithophores significantly alters their sinking rates. For instance, in small *E. huxleyi*, which ranges from 4 to 9 µm in size, calcification changes the sinking speed from 3 cm/d to 30 cm/d - an order of magnitude difference between naked and calcifying organisms [122]. While the literature primarily examines the role of calcification in increasing mass density and sinking rates in coccolithophores [122,148], these organisms can also regulate buoyancy by stopping calcification under extreme light-limited conditions, thereby lowering cell density [149,150]. Under nutrient-limited conditions, such as phosphate or nitrogen deficiency, cell growth is restricted, limiting the use of absorbed light energy for biosynthesis. However, calcification continues despite its energy demand, serving as a mechanism to dissipate excess absorbed light energy that cannot be utilized for growth. This energy dissipation helps protect the cell from potential photodamage [149,151].

Possible sources of hydrodynamic dissipation may also exist within the cell-wall layer, involving the interplay of coccoliths, body scales, and baseplate scales during the calcification process and density adjustment. The coccosphere has multiple roles beyond buoyancy regulation, including mechanical protection, as evidenced by its strong inelastic tolerance to forces [152], protection against photodamage [153], and acceleration of photosynthesis [122]. These diverse roles complicate the assessment of the cost of calcification associated with internal buoyancy-related mechanisms. One approach to assess the role of calcification in buoyancy regulation is to measure the oxygen consumption rate under dark conditions when the phytoplankton is both suspended and sinking. This can be achieved using optical fluorescence oxygen sensors [154] or modern methods that measure extracellular acidification and oxygen consumption rate, providing ATP turnover within the cell [155].

### *5.2.12. Operational Costs: Diatoms*

Buoyancy regulation in diatoms is managed by balancing protoplasm density adjustments and the silicification process in the frustule. The mass density in both the protoplast and frustule can be either increased or decreased relative to water. Optimal buoyancy regulation occurs when these two density adjustments align positively, meaning both compartments either simultaneously increase or simultaneously decrease their mass density relative to water. Conversely, if one compartment increases density while the other decreases it, this misalignment reduces efficiency and increases energy dissipation. The density of protoplasm is changed through the processes like the synthesis of organic osmolytes and trimethyl ammonium [25,62,67], and the active influx of water from the cytosol into the central vacuole [68,83] while the density of frustule is changed through silicification [28,156]. The silicification process generally leads to an increase in frustule density [27]. However, it has been reported that ocean acidification can significantly reduce silica production [157]. Additionally, bacterial attack on diatoms using their hydrolytic enzymes increases the rate of silica dissolution, resulting in a lower mass density of diatoms [158,159].

Active water transport has been proposed as a potential mechanism contributing to buoyancy regulation in large marine phytoplankton. In this mechanism, water is actively transported across the vacuolar membrane from the cytosol into the central vacuole, thereby lowering vacuolar density [68]. The minimum energy required for this process scales with the square of the osmotic difference between the cell interior and the external medium [68]. This scaling arises because water flux increases linearly with the osmotic gradient (for a given membrane hydraulic conductivity), while the energy required to transport water per unit volume against this gradient also increases linearly with the same gradient [68,160]. However, theoretical estimates indicate that, due to unavoidable passive water leakage down the water potential gradient, most of the energy is expended in compensating this backflow rather than generating net transport [68]. For a model diatom (*E. rex*), the net energetic cost is estimated at $7.15 \times 10^{-10}$ W cell$^{-1}$, whereas the gross expenditure reaches $4.84 \times 10^{-8}$ W cell$^{-1}$ (~ $5.8 \times 10^{11}$ ATP s$^{-1}$ cell$^{-1}$), implying substantial dissipation associated with maintaining the gradient against leakage [68]. For this organism the estimated cost of active water transport to generate buoyancy is corresponding to approximately 1.6% of the total cellular biosynthetic energy expenditure [25].

Other possible sources of dissipation might involve cell expansion, which leads to mechanical deformation and the release of elastic energy during mass density changes. This process is also associated with the relative movement of the valves of the frustule, potentially contributing to hydrodynamic dissipation. Given the rigidity of the frustule and the average force required to break it (0.2 – 0.8 mN [161]), any deformation can be energetically costly.

The dynamic sinking behavior observed in large diatoms - characterized by bursts of rapid downward motion lasting a few seconds followed by brief pauses - has been reported in studies by [75,76]. While this unsteady vertical motion enhances nutrient flux, it may also incur a non-negligible energetic cost [83]. Lavoie and Raven have estimated that the minimum energy expenditure for this behavior accounts for 16% of the total energy cost of growth. This is achieved through the modulation of sodium and potassium permeability, interconversion of low-density and high-density organic cations, and alterations in cell expansion rate [83].

The rapidly fluctuating processes are recognized for their high energy dissipation rates in nonequilibrium thermodynamics [162], a factor that should be considered in future analyses. To adjust its internal density, an organism needs to change its mass and volume so that their ratio increases (to move downward) or decreases (to move upward) relative to the density of water. A rapid and sudden change in mass and volume is analogous to the inflation and deflation of a balloon. Classical physics shows that the first attempt to increase the balloon's radius requires much more pressure than subsequent inflations [163,164]. This is due to the resistance of the polymer chains' chemical bonds in the balloon until a threshold stretch is reached [165]. Similarly, in species that exhibit rapid bursts of movement when sinking or rising, the initial sudden effort to increase/decrease the mass density and consequently the speed may dissipate much more energy compared to typical slow adjustments of mass density. This hypothesis could be tested in future experimental studies.

### *5.2.13. Predator and Virus Risk*

In this section, the question is how predation and viral risks *directly* or *indirectly* affect the cost of motility.

(1) The direct cost arises from the increased speed required for evasion, pursuit, or escape, which elevates motility power consumption as described by Stokes' law; in the extreme, this cost may culminate in mortality. Below, we discuss an example illustrating the direct cost of motility in predator-prey encounter rates.

   (A) Meunier *et al.* [166] showed that predator-prey encounters and their cost-benefit outcomes are highly dynamic and depend on multiple interacting traits, including swimming speed, escape behavior, and nutrient status. They show that phosphorus limitation strongly reduces algal swimming speeds, likely through effects on cell structure or overall metabolic rate [166]. Slow prey exhibits low encounter rates but are easy to handle, whereas fast prey is encountered frequently but are more likely to escape or be difficult to capture. Consequently, predators ingest prey with intermediate swimming speeds most efficiently [166]. Overall, prey swimming speed (modulated by nutrient availability) strongly governs predator feeding success.

   (B) In zooplankton, Visser [167] shows that swimming generates an encounter-mediated trade-off: increasing swimming speed enhances prey encounter rates but simultaneously elevates predator encounters, while also incurring energetic costs. Fitness therefore exhibits a non-monotonic dependence on swimming speed, with an optimal intermediate value that decreases as prey concentration increases or predation risk intensifies. In turbulent environments [168], turbulence enhances encounter rates with both prey and predators, reducing the relative advantage of active swimming and, depending on conditions, favoring ambush or suspension feeding, with optimal speeds in some cases approaching zero at high turbulence levels. Overall, motility is advantageous only when the gains in energy intake outweigh both energetic costs and increased predation risk.

(2) Indirect costs happen when the organism must spend part of its energy on defense, such as making toxins or other protective measures. This reduces the energy available for other functions. As a result, even if the swimming cost stays the same, it becomes a larger fraction of the total energy budget. In what follows, we outline three important examples of these indirect costs:

(A) Theoretical models for dinoflagellates show that defense mechanisms such as producing toxins can significantly reduce the rate of cell division especially under nitrogen-starved conditions [169,170]. These toxins are generally produced to deter grazers or competitors, providing a defense mechanism despite the cost to growth. Producing saxitoxin requires at least 132 moles of ATP to synthesize one mole of the substance [170]. The number of moles of saxitoxin per cell can be determined based on the cell's nitrogen content. Assuming $10^{-11}$ g N per cell, with 15% devoted to toxin production [170], and given the molecular formula of saxitoxin ($C_{10}H_{17}N_7O_4$), where the nitrogen content per mole of saxitoxin is $7 \times 14$ g, we calculate: $(0.15 \times 10^{-11}) / (7 \times 14) = 1.5 \times 10^{-14}$ mol of saxitoxin per cell. Multiplying by 132 moles of ATP per mole of saxitoxin gives $2.0 \times 10^{-12}$ mol ATP per cell, which, when converted using Avogadro's number, results in $1.2 \times 10^{12}$ ATP molecules per cell. Under nutrient-limited conditions, this significantly reduces the cell-division rate by more than 20% [170].

(B) Prey can utilize other defense mechanisms, such as the production of blue bioluminescence, which can startle or deter predators and thus reduce the risk of being eaten [171]. This process incurs a high energetic cost due to the chemical reaction between luciferin substrate and luciferase enzyme in the presence of oxygen [172,173] which requires approximately 60 ATP molecules per photon [173,174]. Each bioluminescent flash can last 0.1 to 0.5 s and can involve the production of $10^7$ to $10^{11}$ photons per second [175], equating to a minimum of $60 \times 0.1 \times 10^7 = 6 \times 10^7$ to a maximum of $60 \times 0.5 \times 10^{11} = 3 \times 10^{12}$ ATP molecules per flash.

(C) Phytoplankton may be infected by viruses, and such infections impose both direct and indirect costs beyond increased mortality. These include the energetic cost of cellular resistance and the cost associated with infection-induced changes in sinking or swimming behavior. Resistance to viral infection (defined as the inability of viruses to successfully infect and replicate within host cells [176]) can arise through multiple mechanisms, including modification or loss of cell surface receptors [177], among others. However, such modifications may compromise cellular functions involved in nutrient uptake and metabolism, thereby reducing resource acquisition efficiency and imposing physiological and energetic costs that manifest as reduced competitive fitness [176,177].

Viral infection can also alter the vertical motility of phytoplankton, often increasing sinking speed, as reported for *Heterosigma akashiwo* [178] and *Aureococcus anophagefferens* [179]. This behavior has been interpreted as an adaptive, population-level response that may limit viral spread to conspecifics [180]. However, the underlying biophysical mechanism remains unresolved. Several hypotheses have been proposed: for example, enhanced synthesis and accumulation of dense viral proteins within infected cells may increase cellular mass density, thereby accelerating sinking [179]. Alternatively, faster descent may transport infected cells to deeper, low-light environments where viral replication is less efficient, potentially allowing the host to suppress the infection or transition into a dormant, cyst-like state [179].

Studies on flagellated bacteria have shown that, while swimming may increase susceptibility to viral infection, hydrodynamic interactions mitigate this risk around the cell body but exacerbate it near the flagellar bundle. This is a consequence of differences in fluid flow dynamics around the flagella compared to those around the cell body [181]. Experiments with ciliates fed various viruses, including mammalian viruses and bacteriophages, indicate that the presence of certain phages progressively increases the swimming speed of ciliates. This increase in swimming speed is interpreted as a behavioral response, suggesting that *Tetrahymena pyriformis* senses these viruses and recognizes them as potential food, similar to its response to nonviral prey [182]. The precise costs and benefits associated with viral infection on the swimming or sinking behavior of phytoplankton remain unknown.

Table 2 provides a summary of all the motility costs discussed in this paper.

*Table 2: Summary of motility costs for various swimmers and buoyancy-regulating organisms, including flagellated phytoplankton, cyanobacteria, diatoms, and coccolithophores.*

<table>
<tr><td rowspan="3">Phytoplankton Motility Costs</td><td rowspan="3"><u>Swimming Machineries</u></td><td>Construction costs</td><td colspan="2">- Synthesis and remodeling of flagellar proteins and membrane lipids, including costs associated with flagellar loss and length fluctuations.<br>- Crystalline surface layer, SwmA glycoproteins, SwmB polypeptides, and peptidoglycan-associated proteins.<br>- Type IV pilus (T4P) machines.</td></tr>
<tr><td rowspan="2">Operational costs</td><td>External efficiency</td><td>- Minimum hydrodynamic power required to overcome viscous drag.<br>- Geometric shape of the cell body.<br>- Flagellum-to-body length ratio, flagellar slenderness, number, geometry, angle, and arrangement of flagella.</td></tr>
<tr><td>Internal efficiency</td><td>- Elastic bending of the axoneme.<br>- Sliding of microtubules.<br>- Internal viscosity.<br>- Frictional/mechanical dissipation as motor proteins move beneath and interact with the S-layer to generate surface waves.<br>- Proton leak.<br>- Retraction process.<br>- Jumping behavior.</td></tr>
</table>

<table>
<tr>
<td rowspan="3"></td>
<td rowspan="3"><u>Buoyancy Regulating Machineries</u></td>
<td>Construction costs</td>
<td colspan="2">- Synthesis of gas vesicles, carbohydrates, organic osmolytes, and DMSP.<br>- Formation of frustules, coccolith layers, and silica cell walls.<br>- Polysaccharide synthesis and storage.<br>- Channels mediating water influx/efflux in vacuoles.<br>- Proteins responsible for ion exchange of heavy and light ions in vacuoles.</td>
</tr>
<tr>
<td rowspan="2">Operational costs</td>
<td>External efficiency</td>
<td>- Geometric shape of the cell body.</td>
</tr>
<tr>
<td>Internal efficiency</td>
<td>- Minimum mechanical expenditure associated with active shape and density adjustment.<br>- Active water transport across the vacuolar membrane.<br>- Energy required to establish and maintain osmotic and water-potential gradients.<br>- Compensation for passive water leakage or backflow.<br>- Ratio of gas space to protein wall volume in gas vesicles.<br>- Turgor or hydrostatic pressure.<br>- Elastic modulus of gas vesicles.<br>- Density change alignment between protoplasm and frustule.<br>- Cell expansion, hydrodynamic dissipation, and relative movement of frustule valves.<br>- Costs associated with rapid bursts of movement, including modulation of sodium and potassium permeability.<br>- Continuous calcification under nutrient-limited conditions.</td>
</tr>
<tr>
<td>Other Costs (for both swimmers & sinkers)</td>
<td colspan="4">Direct costs:<br>- Increased speed due to viruses/predator risk, and turbulence.<br>- Increased predator-encounter risk and viral-contact or infection risk caused by active motility.<br>- Mortality risk resulting from predator or viral encounters.<br>- Environmental high local viscosity gradients.<br>- Navigation costs.<br>- Reorientation and change in motion direction requiring dynamic body shape adjustments, mass density modifications, and alterations in center of mass and buoyancy, which impact both external and internal motility efficiency.<br>- Chaotic path.<br><br>Indirect costs:<br>- Defense mechanisms (e.g. production of toxins and bioluminescent flashes, and resistance to viral infection), along with nutrient and light limitations during migration, reduce total metabolic power, leading to relatively higher motility costs.</td>
</tr>
</table>

### *5.3. Motility Cost to Total Metabolic Rate Ratio*

To determine what percentage of the total metabolic rate of the cell is accounted for by the motility mechanisms, one requires data on the total metabolic rate of the studied species. This task is more straightforward for heterotrophic protists under growing or starved conditions, as it can be obtained through the rate of oxygen consumption [183–187]. However, this poses a greater challenge for phytoplankton cells, whether they are photoautotrophs or mixotrophs, because these microorganisms are both producers of oxygen in their chloroplasts and consumers of oxygen in their mitochondria and cytoplasm [188]. ATP is widely regarded as the universal energy currency of most living systems [189], with both constructional and operational costs typically expressed in terms of ATP molecules [49,190].

We evaluate the potential costs associated with phytoplankton motility using the dinoflagellate *Prorocentrum mariae-lebouriae* as an example. This species is currently considered a synonym of *Prorocentrum minimum* (Pavillard) J.Schiller [191]. *Prorocentrum minimum* is a geographically widespread, bloom-forming dinoflagellate linked to harmful algal blooms, ecosystem damage, and potential shellfish toxicity, often associated with coastal eutrophication [192]. This dinoflagellate measures 14.8 µm in length, 14.8 µm in width, 7.3 µm in depth, and 11.7 µm in equivalent spherical diameter, with an oblate shape and an aspect ratio of 0.49 [193]. It has a volume ($V$) of 846 µm$^3$ and a mass density of 1064 kg/m$^3$, resulting in a total body mass of approximately 0.9 ng and a swimming speed of 170 µm/s at 20ºC [193].

The construction cost of flagella ($C_\mathrm{f}$ [#ATP]) can be estimated using a fitting equation provided by [24]:

$$C_\mathrm{f} = 2.6 \times 10^{10} V^{0.31} \quad (9).$$

This yields a cost of approximately $2.1 \times 10^{11}$ ATP molecules, or 17.4 nJ, for the dinoflagellate under study. As described in previous sections, the length of the flagella can vary by up to ±20% due to fluctuations in the removal and addition of amino acids and proteins. Consequently, the construction cost ranges approximately from 13.9 nJ to 20.8 nJ. The construction cost of flagella should be evaluated in relation to the construction cost of the entire cell. We assume that the cell's construction cost is determined by the total energy content of all its constituent proteins, carbohydrates, and lipids. A dinoflagellate with a volume of 846 µm$^3$ has a total chemical energy content $E\left[\frac{\mathrm{nJ}}{\mathrm{cell}}\right]$ (comprising proteins, carbohydrates, and lipids) of $5.3 \times 10^3$ nJ/cell. This value is derived using the equation developed for dinoflagellates [194]:

$$\log_{10} E = 2.88 + 0.91 \log_{10}\left(\frac{V}{100}\right) \quad (10).$$

Thus, the construction cost of the flagella ranges from 0.26% to 0.39% of the total construction cost of the *P. minimum* body.

Next, we estimate the operational cost as a fraction of the organism's total metabolic rate. Based on Stokes' law (Equation 1), the minimum power required for the organism to tow its spherical body can be estimated as $P = 6\pi\eta r v^2$ where the viscosity of seawater is $\eta \approx 1.09 \times 10^{-3}$ kg m$^{-1}$ s$^{-1}$ at 20°C. However, as discussed in this review, several corrections to this equation must be considered.

1. For external efficiency related to the physical properties of flagella, such as the flagellum-to-body length ratio, and slenderness of the flagellum, we consider the average efficiency reported for green algae [20,21] to be 1%. This indicates that for every 100 units of energy, 99 units are dissipated, meaning the organism must expend 99 times more energy to compensate for the energy required for swimming. Consequently, the swimming minimum power consumption estimated by Stokes' law should be multiplied by 100.
2. In the above calculation of external efficiency, the organism's body is approximated as a rigid sphere [20,21]. According to Figure 3 in [17], the shape-related efficiency of a spherical microswimmer is approximately 1/3 (i.e., 0.33). In contrast, the dinoflagellate considered here has an aspect ratio of 0.49. By fitting a function to the data presented in Figure 3 of [17], we estimate an efficiency of 0.16 for *P. minimum*. This indicates that the efficiency of this oblate-shaped dinoflagellate is roughly twofold lower than that of a spherical counterpart (0.33/0.16 ≈ 2.1). Therefore, the operational cost should be further multiplied by approximately 2.1.
3. Next, we consider the elastic cost of bending the axoneme, which is associated with swimming internal efficiency [22,23]. As reported by [22], the chemomechanical efficiency (referred to in our article as internal efficiency) of sea urchin sperm ranges from 34% to 60%, depending on the viscosity. We use a representative value of 50%, meaning that the operational cost should be multiplied by 2.
4. We do not include the cost of jumping, as it varies between species (and not all species exhibit jumping behavior). Additionally, accurate estimation requires knowledge of the frequency and duration of jumps.
5. In this estimation, we assumed the dinoflagellate relies solely on swimming without buoyancy regulation. However, since dinoflagellates can perform both, if they do so simultaneously, the molecular costs and internal efficiency of mass density adjustment would factor into the above calculations, potentially increasing the overall motility cost.

Therefore, the power consumed during swimming can be calculated as follows:

$$
\begin{aligned}
P_{\text{tot}} &= (6\pi\eta r v^2 \times 100 \times 2.1 \times 2)\ \text{pW} \\
&= 6\pi \times 0.00109[\text{kg.m}^{-1}\text{s}^{-1}] \times 5.85\times10^{-6}\,[\text{m}] \times (170\times10^{-6})^2\,[\text{m}^2\text{s}^{-2}] \times 100 \times 2.1 \times 2 \times 10^{12} \\
&\approx 1.5\ \text{pW}. \qquad (11).
\end{aligned}
$$

To assess whether this operational cost is significant, it should be compared to the organism's total metabolic rate - that is, its overall rate of ATP production.

A precise assessment of the total ATP production rate in phytoplankton requires knowledge of the respiration rate [195], mitochondrial proton leak [134,196], ATP yield per oxygen molecule consumed in mitochondria [197], photosynthesis rate [198,199], proton leak across the thylakoid membrane of the chloroplast [200,201], and the total ATP yield per oxygen molecule produced in the chloroplast, accounting for non-cyclic, cyclic, and pseudo-cyclic electron transport pathways [202,203]. However, a detailed assessment is beyond the scope of the present work (see [204] for a comprehensive analysis). Therefore, we provide a rough estimate to illustrate how the total available power content per organism can be approximated.

One might question that ATP produced in chloroplasts is entirely consumed by carbon fixation and other intra-chloroplast processes, so why should the cost of motility be evaluated relative to the total energy produced by both mitochondria and chloroplasts? Mitochondria and chloroplasts function in a tightly interconnected manner. For example, in diatoms, chloroplasts generate oxygen during photosynthesis, which can be utilized by mitochondria for respiration in addition to oxygen absorbed from the surrounding environment [205]. In turn, mitochondria supplement chloroplasts by producing and supplying ATP when photosynthetic ATP production is insufficient to meet the chloroplasts' energy demands [205]. This energetic interplay is supported by a dynamic exchange of reducing equivalents (electron carriers) between the two organelles, underscoring their close metabolic interdependence and coordinated functioning [205]. Given this interdependence, we propose that motility cost should be evaluated relative to the combined energy production from both mitochondria and chloroplasts.

The respiration rate in phytoplankton - based on a general relationship not specific to dinoflagellates - is scaled with cell volume according to the equation [206]:

$$\log R = 0.9 \log V - 4.1 \quad (12),$$

where $R$ is the respiration rate in pmol $O_2$ cell$^{-1}$ h$^{-1}$, $V$ is the cell volume in μm$^3$, and the equation applies at 18°C under exponential growth. Assuming that each mL of $O_2$ yields approximately 20 J of energy [183], the respiration rate for *P. minimum*, with a cell volume of 846 μm$^3$, corresponds to 4.27 pW. Adjusting this to 20°C using a $Q_{10}$ of 2.5 increases the value to 5.13 pW. In actively growing dinoflagellates, the ratio of respiration to photosynthesis spans approximately 0.08–0.6 [198,199,206]. Based on this range, and assuming that molecular oxygen generated in chloroplasts yields the same amount of ATP as molecular oxygen consumed by mitochondria, the corresponding photosynthetic energy production rate is estimated to lie between 8.55 pW (5.13/0.6) and 64.125 pW (5.13/0.08). Adding respiration and photosynthesis therefore results in a total metabolic power between 13.68 pW (8.55 + 5.13 pW) and 69.25 pW (64.12 + 5.13 pW) for *P. minimum* at 20 °C.

Thus, the previously estimated swimming operational cost of 1.5 pW can account for anywhere between 2.2% and 11% of the organism's total metabolic rate - a substantial proportion. This relative cost reflects the intrinsic energetic burden associated with active relocation in this species. This cost is likely to increase under nutrient or light limitation during migration.

To generalize the swimming-related operational costs estimated above across a broad range of dinoflagellates, we use swimming speed and cell volume data reported in [193,207–209]. A limitation of this dataset is that cell shape (e.g., oblate, prolate, or spherical) and aspect ratio are not provided, except in one study [193]. We therefore repeat the calculations assuming spherical geometry to obtain a first-order estimate of the range of operational energetic costs, while bearing in mind that these costs may be lower or higher due to the increased or reduced efficiencies associated with organismal shape. Our estimates (see Supplementary Data) indicate that for *Peridinium crassipes*, with a volume of $2.04\times10^{5}$ μm$^3$ and a swimming speed of 100 μm s$^{-1}$, the swimming operational cost is as low as ~0.02 % to 0.08% of the total metabolic rate. In contrast, for *Katodinium rotundatum*, with a volume of $5.3\times10^2$ μm$^3$ and a swimming speed of 430 μm s$^{-1}$, the swimming cost is estimated to range from ~8.40% to 42.5% of the total metabolic rate.

Dinoflagellates actively respond to turbulent conditions using various strategies, including density adjustment, shape modification, and changes in orientation [102]. These dynamic changes affect swimming power consumption, as they are directly related to both speed and the swimming internal and external efficiencies. The density changes in dinoflagellates are proposed to be mediated by the cell nucleus, with a mass density of 1300 kg/m$^3$, and by 15-20 chloroplasts, each with a mass density of 1150 kg/m$^3$ [102]. The actual costs associated with these rapid changes under turbulence remain unknown, representing a gap in our knowledge. However, it has been observed that the rate of stress, measured as the rate of production of the nitric oxide (NO) free radical under turbulent conditions, increases [102]. Estimating the energetic cost of NO production and its accumulation in the cell constitutes the second gap for future studies, even though the mechanisms involved in the production of this radical in living systems are known [210].

Because motility speeds are not independent of organism size, the scaling between size and speed introduces an additional constraint on the fraction of metabolic energy that can be allocated to movement; a synthesis of reported size–speed relationships is provided in Appendix.

Providing an explicit quantitative example of the energetic costs of buoyancy regulation remains a subject for future work. Although the energetics of the underlying molecular mechanisms have been studied extensively [25,67,68,83], the efficiencies of these mechanisms have not yet been examined.

## 6. Temperature-Dependent Costs of Motility

The projections regarding the future of climate change and global warming raise several concerns, one of which revolves around the fate of marine phytoplankton and their role in carbon cycling within the ocean [211–219]. Establishing a mechanistic model that links motility costs to temperature variations appears complicated, given the myriad of effective parameters discussed in previous sections.

In a recent study [220] on phytoplankton behavior, researchers found that warming, rather than nutrient availability, promoted the selection for motility in diatoms, a pattern that may be related to temperature-dependent changes in the mechanism of locomotion, as increased temperature has been shown to decrease cytoplasmic viscosity and allow motile diatoms to move faster. Additionally, warming intensifies stratification by creating stronger temperature gradients in the water column, which limits vertical mixing and leads to nutrient-depleted surface layers. Under these conditions, motility becomes a crucial adaptive trait, enabling organisms to actively navigate between nutrient-rich deeper layers and light-rich surface waters. According to [220], seasonal temperature variations did not significantly influence motility in benthic diatoms, suggesting that the effects of sustained warming differ fundamentally from natural seasonal changes.

The effect of ocean acidification on motility in aquatic organisms appears to be context-dependent, varying by species, cell type, and duration of exposure. In the harmful algal species *Heterosigma akashiwo*, both short-term and generational-scale exposure to elevated $CO_2$ (low pH) led to enhanced motility - particularly increased downward swimming speed. In contrast, studies on flagellated algal species (*Microglena sp.*, *Dunaliella salina*, and *Chlamydomonas reinhardtii*) [221] demonstrated that acidification impairs motility. Notably, the long-term (5-year) exposure study revealed a conserved reduction in motility across diverse taxa, suggesting that elevated $CO_2$ may broadly compromise the flagellar machinery across eukaryotic lineages [221]. On top of these temperature and $CO_2$ effects, evolutionary and community composition changes can further influence motility patterns over time.

Stokes' law (Equation (1)) provides the idea that estimating the temperature-dependent cost of motility requires knowledge of how viscosity, cell size, and motility speed vary with temperature. When combined with the temperature dependence of the organism's metabolic rate, these factors can be used to estimate the relative cost of motility across temperature gradients. However, even this simplified scenario is inherently complex, and the actual situation is even more so. We begin by focusing on the temperature dependence of viscosity, cell size, motility speed, and metabolic rate. We then address other indirect effects of temperature on the cost of motility.

Temperature affects both the viscosity and density of seawater, which in turn influences hydrodynamic forces. As temperature increases, water becomes less viscous and less dense, reducing the resistance experienced by objects moving through it. These changes modify the Reynolds number, leading to variations in drag and inertial forces [222].

Research on the temperature dependence of organism size shows that increasing temperature leads to a systematic reduction in body size, driven primarily by within-species phenotypic size decreases (the temperature–size rule), with individuals becoming smaller (often approximately linearly) with warming [223,224]. Additional reductions in mean community size can arise from temperature-driven shifts toward smaller-bodied species, but these represent a distinct, community-level mechanism [223,224]. Temperature is indeed recognized as a major factor influencing size structure, both directly and indirectly [223]. Although no definitive general theory has been established for the temperature-size rule, several hypotheses have been proposed to explain this phenomenon:

(1) First, as temperature rises, the availability of essential resources such as carbon dioxide and dissolved oxygen decreases, while metabolic demand increases. Smaller cells can more effectively absorb and utilize these limited resources due to their higher surface area-to-volume ratio, thereby compensating for the reduced resource availability [224]. Contrary to this hypothesis, some bacteria do not depend on gases for survival. Additionally, nutrient availability is not directly temperature-dependent, as it varies by geographical location - some regions become depleted around 20°C, while others do so near 0°C [225].
(2) Second, smaller cell size may facilitate faster reproduction, enhancing survival and competitive advantage [224]. However, this raises the question: why don't organisms remain small in size permanently? Contrary to these propositions, Maranon *et al.* [226]

have reported that there is no overarching law elucidating the impact of temperature on phytoplankton size structure. They emphasize that resource availability serves as a major factor influencing size structure. An example that contradicts Atkinson's size-temperature rule is the finding that *Thalassiosira pseudonana* diatoms increase in size with rising temperatures in evolved lineages after 300 generations of adaptation [227]. This size increase is hypothesized to enhance lipid content, serving as a strategy to mitigate oxidative stress induced by high temperatures [228].

Beveridge *et al.* [229] introduced a model to describe different scenarios involving the temperature dependency of protists' swimming speed, influenced by the interplay between temperature-dependent viscous drag and metabolic rate. The formula they have derived for swimming speed integrates Stokes' law, mass-specific metabolic rate, and the temperature dependencies of viscosity and size, expressed as follows:

$$v = \sqrt{\frac{a_0 i_0 [M(T)]^{3/4} \mathrm{e}^{-E/k_B T}}{6\pi\eta(T) r(T)}} \quad (13),$$

where $a_0$ represents the proportion of metabolic rate allocated to swimming, $i_0$ is a normalization constant with the units $kg^{1/4}.m^2.s^{-3}$, $M(T)$ is the temperature dependence of body mass, derived from the assumption of a 2.5% decrease in cell size per 1°C temperature increase in protists [224], $k_B$ is the Boltzmann constant, and $E$ is the activation energy. Their findings suggest that swimming speed escalates with rising temperatures [229], aligning with other research exploring temperature-dependent sinking speeds [139,230]. However, Kamykowski and McCollum's investigation of dinoflagellates' temperature-acclimatized swimming speed across 5–38°C under stable nutrient and light conditions reveals a trend where speed increases with temperature until reaching a threshold, beyond which it declines [231]. They report that the temperature range supporting swimming ability is broader than the range permitting growth [231]. In their study, the maximum swimming speed is observed in cells approximately 35 µm in length. This decline in speed at higher temperatures and larger sizes can be attributed to physiological limitations and inefficiencies in motility machinery - factors not accounted for by the Arrhenius equation.

To capture how total metabolic rate (i.e., ATP production rate) responds to temperature, one should consider several underlying parameters at play. Respiration rate, for instance, typically follows a Q10 between 2 and 3, averaging around 2.5 [232]. Proton leakage increases with temperature, causing a portion of the metabolic energy to be lost as heat, as demonstrated in mitochondria from the eurythermal bivalve *Mya arenaria* [233]. The ATP yield per molecular oxygen consumed in mitochondria also shows temperature dependence, as seen in intact mitochondria of *Arabidopsis thaliana* [234]. This is because the electron transport chain (ETC) is less sensitive to temperature changes than ATP synthase [234]. Temperature also influences both the catalytic activity and abundance of ATP synthase in plant mitochondria [235]. Furthermore, the respiration-to-photosynthesis ratio shifts with temperature due to differing thermal sensitivities of respiration and photosynthesis, as observed in *Chlorella vulgaris* [236]. Under nutrient-limited conditions, the activation energy of photosynthesis can be up to three times that of respiration, as shown in *Emiliania huxleyi*, *Skeletonema costatum*, and *Synechococcus* species [237].

In the following, we present two examples illustrating how temperature influences the motility machinery (in this case, flagella) and its efficiency, concluding this section by highlighting the challenges involved in modeling the temperature-dependent costs of motility:

(1) Assembly and function of flagella [238], the ATP-hydrolysis rate by molecular motors involved in motility [239], and the internal viscosity and density, are all temperature-dependent, influencing internal efficiency in swimming or sinking. Moreover, temperature-dependent changes in surface tension [240,241], elasticity [241–243], and permeability [244] of the cell membrane result in alterations in cell shape and size, thereby affecting external efficiency in swimming or sinking.

(2) Researchers studied how temperature affects the flagella of the algal species *Chlamydomonas reinhardtii* [238]. They found that when the temperature shifts from 20°C to 32°C, the flagella respond differently depending on the specific mutant being observed. This range of responses includes flagella becoming broken down, detached, or malfunctioning, all of which incur energetic costs [238]. Beyond these structural effects, the beat frequency of eukaryotic cilia and the rotation of bacterial flagella are also temperature dependent, reflecting the combined effects of temperature on molecular motor activity and the viscosities of both the cell and the surrounding fluid [245].

## 7. Conclusion

This work presents a comprehensive re-evaluation of phytoplankton motility, challenging the traditional dichotomy between "active" swimming and "passive" sinking. By integrating biological, physical, and ecological evidence, we propose that both swimming and buoyancy-regulated movement can involve substantial energy costs. The overarching picture that emerges is that motility in phytoplankton is not a binary trait but a continuum of energetically demanding strategies shaped by physiological and environmental trade-offs. These costs, often underestimated or overlooked, must be balanced against survival benefits such as nutrient acquisition, predator avoidance, and successful migration through heterogeneous marine environments. This perspective reframes our understanding of phytoplankton behavior and physiology, offering an energetic basis for explaining ecological patterns such as the paradox of slower-growing, yet more motile, species like dinoflagellates.

We argue that, for swimmers, the efficiency of swimming machinery has been investigated by biophysicists but largely overlooked in marine and freshwater sciences, resulting in underestimation of swimming energetic costs. Conversely, although the molecular mechanisms of buoyancy regulation have been extensively studied by marine and freshwater scientists, the efficiency of buoyancy-regulating machinery has been entirely overlooked by both biophysicists and biologists and therefore warrants focused investigation in future studies.

We are still some distance away from achieving a comprehensive cost-benefit analysis of phytoplankton swimming and sinking behaviors, with numerous uncertainties remaining. Some less-explored topics encompass the cost-benefit analysis of dynamic strategies employed in response to migratory or turbulent conditions, such as sudden bursts of movement, alterations in shape and orientation, fluctuations in mass density, formation of chains, subdivision into subpopulations, viral infections, depth-dependent environmental and physiological changes, temperature dependent motility variations, and viscosity gradients.

We synthesized diverse aspects of the costs and benefits of phytoplankton motility, ranging from the level of a single cell to the scale of the entire ocean. Our goal has been to achieve an integrative understanding of this phenomenon by utilizing the current knowledge available in the literature and proposing new avenues for future investigations. We conducted a thorough examination of the motility costs of a model dinoflagellate, estimating these costs relative to the organism's available energy resources.

We find that the efficiency and construction costs associated with sinking/rising-related machinery, as well as the swimming mechanisms of surface-distorted cyanobacteria, remain largely unexplored. Phytoplankton sinkers/risers are active, motile microorganisms with invisible buoyancy regulation mechanisms. We investigated the construction costs and potential sources of internal energy dissipation related to motility in diatoms, cyanobacteria, and coccolithophores. This can aid biophysicists in developing equations to model the internal efficiency of buoyancy-regulating phytoplankton.

The physics underlying the shape-related external efficiency of sinkers remains an open problem. For example, as discussed in this review, theoretical derivations of the drag coefficients for perfect-slip and no-slip rigid bodies in swimmers enable the quantification of shape-related efficiency and explicitly describe how this efficiency varies from oblate to prolate geometries. In contrast, for sinking organisms, there is currently no comparable theoretical framework that explains why tear-drop-shaped bodies sink faster than spherical ones, or why spherical shapes outperform other geometries. Instead, most findings for sinkers remain largely empirical, lacking a unifying and physically grounded theoretical explanation.

Oblate-shaped phytoplankton swimmers with an aspect ratio of ~0.01 exhibit remarkably high swimming external energy dissipation, up to $10^6$ times that of prolate-shaped microswimmers with an aspect ratio close to 100, making swimming a highly inefficient strategy. Why do these species possess swimming abilities despite such inefficiencies? For example, in bacteria, flagella serve multiple functions beyond swimming, including mechanosensing [246], wetness sensing [247], adhesion [248], and contributing to virulence [248].

## 8. Acknowledgment

The authors thank Professor Zoe V. Finkel (Dalhousie University), Dr. M.M. Amirian (Dalhousie University), Professor S. Mukherji (Washington University in St. Louis), and Professor John Cullen (Dalhousie University) for helpful discussions, and the editor and anonymous reviewers for their constructive comments, which improved the manuscript. This work was supported by the Simons Collaboration on Computational Biogeochemical Modeling of Marine Ecosystems (CBIOMES) (Grant ID: 549935 to AJI) and by grants to ML from the National Institutes of Health

(2R35GM122566), the National Science Foundation (DBI-2119963, DEB-1927159), the Simons Foundation (735927), and the Moore Foundation (12186).

## 9. Conflicts of Interest

No conflicts of interest to declare.

## 10. Appendix

Table S1 summarizes idealized point-force models used in the literature to characterize swimming-induced flow fields and associated feeding currents in phytoplankton.

*Table S1: The flow velocity equations modeled for various phyla of flagellated phytoplankton [31], are generated from time-averaged point forces with magnitude F applied at a source point. These equations describe the flow velocity at a field point (x, y, z) in Cartesian coordinates, where* $d = \sqrt{x^2 + y^2 + z^2}$ *represents the distance between the origin and the field point. The chrysophyte studied in [31] is a sessile cell that generates a feeding current using a single beating flagellum (see Movie S2 in [31]). This beating flagellum can be modeled as a point force acting at the origin, directed along the z-axis in 3D space, with x, y, and z representing the spatial coordinates of the beating flagellum. The resulting velocity field resembles that of a stokeslet, which is a fundamental solution to the Stokes equation in free space (without boundaries) (Pozrikidis, 1992). In the velocity field equation,* $\eta$ *represents the viscosity of the fluid. The raphidophyte studied in [31] propels itself by generating force at the front of its body through the motion of its flagella (see Movie S1 in [31]). In these organisms, the flagellar beat is typically oriented so that the flagellum pulls the organism forward through the surrounding fluid, rather than pushing it from behind (meaning the force applied by the flagellum is opposite to the force applied by the cell body). For the derivation of the equation, refer to the supporting information in [56]. The haptophyte and chlorophyte studied in [31] are assumed to propel themselves using a three-stokeslet quadrupole point force model, where two or more flagella generate, mainly, lateral propulsive forces of equal magnitude, pushing the flow backward, while the opposing drag force of the moving cell body balances the sum of these propulsive forces (see Movie S4 in [31]). The flow velocity field equation is developed in [57].*

| Phylum | Flagellar arrangement | Point force model | Equation | Ref. |
|---|---|---|---|---|
| Chrysophyte | One beating flagellum | Stokeslet: A single point force | $v = \frac{F}{8\pi\eta} \frac{\sqrt{x^2 + y^2 + 4z^2}}{d^2}$ | Supplementary information in [31]. |
| Raphidophyte | One beating flagellum + Opposing force of the cell body | Puller stresslet: Two opposing equal point forces at positions $(x = 0, y = 0, z = \pm b)$ | $v = \frac{Fb}{4\pi\eta} \frac{\lvert x^2 + y^2 - 2z^2 \rvert}{d^4}$ | |
| Haptophyte Chlorophyte | One or more lateral flagella + Opposing force of the cell body | Three-stokeslet quadrupole: Two opposing equal point forces at positions $(x = \pm a, y = 0, z = 0)$ and a third, doubled, point force applied at the origin | $v = \frac{Fa^2}{8\pi\eta} f(d)\,*$ | |

* $f(d) = \left(\sqrt{4x^4(x^2 + 21z^2) - 3x^2(y^2 + z^2)(y^2 + 13z^2) + (y^2 + z^2)^2(y^2 + 16z^2)}\right)/d^6$

## Coupling of Phytoplankton Size and Motility Speed

Using Stokes' law to estimate the operational cost of motility raises the question of whether the parameters in the equation are interrelated. One important potential coupling is whether swimming or sinking speed varies with organism size. Recent studies [24,29,31,250,251] consistently suggest that swimming speed of unicellular eukaryotes shows either no dependency on cell volume or a weak sub-linear dependence. However, some older references (e.g. [193]) propose that intermediate-sized dinoflagellates display the highest swimming speeds.

A challenge encountered in the analyzed data across the literature is the limited sample size, which often fails to encompass the entire spectrum of volume and speed ranges. Additionally, data collected from different studies covering a wide array of sizes and speeds may vary in terms of temperature, physiological conditions, and other influencing parameters. Hence, the initial approach to addressing this issue involves acquiring a comprehensive and uniform dataset that spans across all size and speed ranges ideally normalized to temperature. A similar situation applies to the sinking speed of species. Generally, it is reported that the sinking rate increases linearly with cell size [74,252], but this relationship is not considered universal due to the high uncertainties between the scattered empirical data and the theoretical models.

One strategy for dealing with highly scattered data involves identifying nonlinear patterns through smoothing techniques. One such method is Locally Weighted Scatterplot Smoothing (LOWESS) [253], which segments the dataset into subregions, conducts linear regressions within each subregion, and then connects them using a cubic function. Here, we apply this approach to analyze the swimming speed versus size data of 111 flagellates and 78 ciliates collected by [29] from different references, along with data on sinking speed of marine particles and living species versus size obtained from 5650 sinkers collected in [252]. The data on flagellates and ciliates follow a log-normal distribution [29], whereas the log-transformed data of sinking particles and species exhibit a bimodal distribution with negative skewness according to our analysis. The result is shown in Figure S1.

When should these nonlinearities be considered in data analysis? Nonlinear data analysis is necessary when nonlinear regression is statistically significant with a considerable confidence interval and when the observed nonlinearity, along with its associated variations and noise, has a meaningful physiological interpretation.

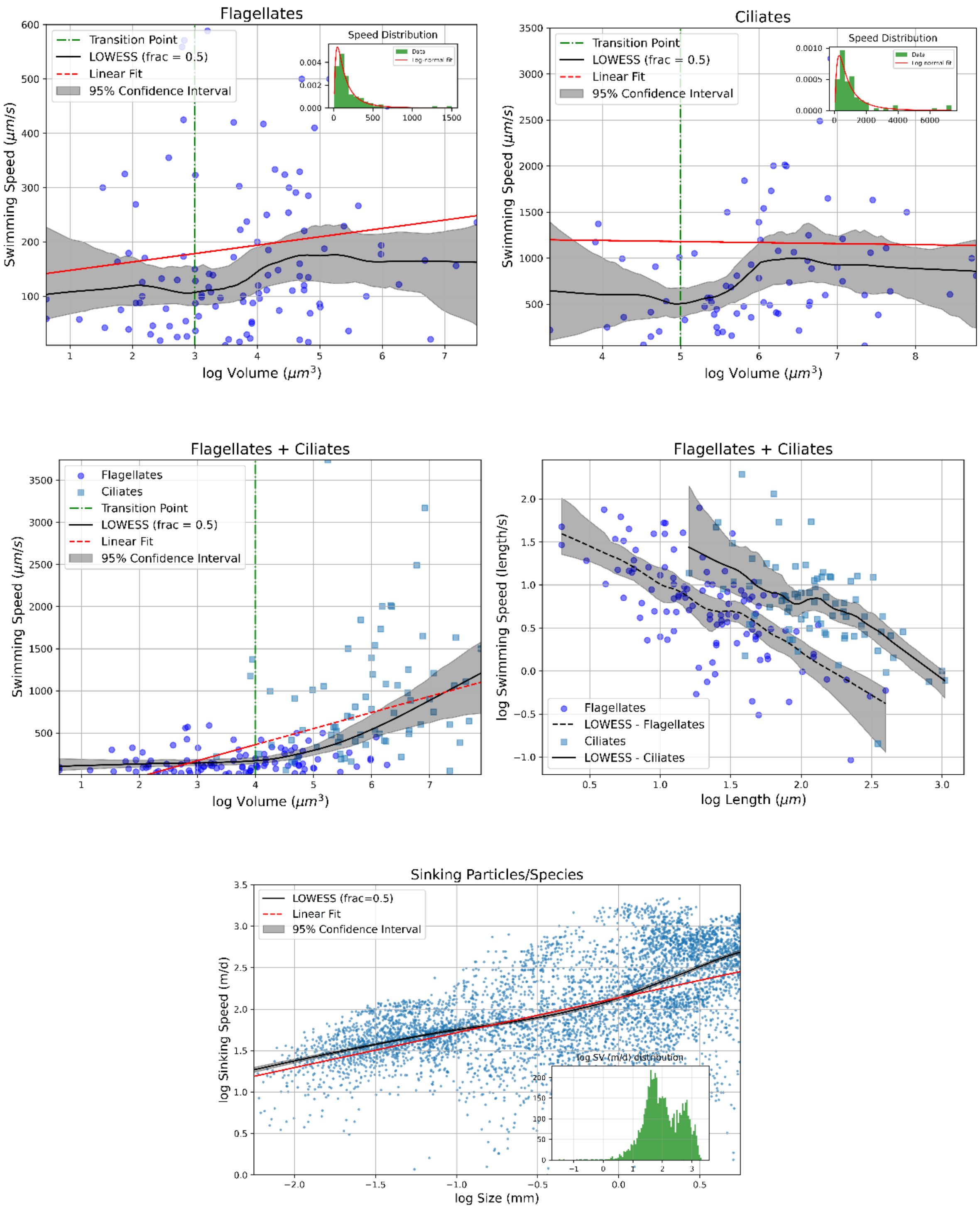

*Figure S1: LOWESS regression curves depicting the semi-log volume[μm$^3$]-speed[μm/s] relationship for flagellates (top-left), ciliates (top-right), a combination of flagellates and ciliates (middle-left), log-log length[μm]-speed[length/s] relationship for both flagellates and ciliates (middle-right), and the log-log size[mm]-speed[m/d] plot of sinking species/particles (bottom).*

In Figure S1, none of the nonlinearities are significant except in the middle-left panel, which shows a transition in speed across species from flagellates to ciliates. In Figure S1, the semi-log plots in the top row - flagellates (upper left) and ciliates (upper right) - do not demonstrate a statistically significant nonlinear or linear pattern. However, the steepest slopes of the LOWESS curves in flagellates and ciliates are considerably greater than the slopes of the corresponding linear regressions specifically in ciliates. Combining data for flagellates and ciliates (middle-left) reveals a statistically significant increase in swimming speed (μm/s) for log volumes ($\mu m^3$) greater than 4 (vertical green line), attributable to taxonomic differences. This suggests an average increase in speed during the transition from flagellates to ciliates. The middle-right figure represents the log-log plot of swimming speed (body-lengths per second) versus body length (μm) for both flagellates and ciliates. As shown, swimming speed relative to body length decreases with increasing cell size in both groups, exhibiting similar slopes. However, this analysis represents speed-to-length ratio versus length (a spurious correlation of ratios), meaning that since both axes include length, it may introduce a statistical artifact rather than capturing a true biological trend. Given the lack of clear linear or nonlinear patterns in the top-left and top-right figures, along with substantial noise in the data, the observed decreasing trend and similar slopes between flagellates and ciliates in the middle-right figure could be an artifact of division bias. Therefore, this trend requires further investigation in future studies. In sinking species/particles (bottom, log-log plot), the LOWESS curve diverges from linear regression for log sizes (in mm) exceeding zero with a high confidence interval. Based on this analysis, we cannot yet confirm or refute the hypothesis of a relationship between size and speed in phytoplankton swimmers and sinkers.

We also analyzed the dataset compiled by [254], which includes measurements of body length and swimming speed for bacteria, archaea, flagellates, ciliates, and spermatozoa. Our findings align with those discussed in Figure S1. The only distinct pattern is observed in spermatozoa, which show a slight negative correlation between swimming speed and body length, suggesting that larger sperm tend to swim more slowly. However, this correlation is weak.

## 11. References

[1] Finkel ZV, Beardall J, Flynn KJ, Quigg A, Rees TAV, Raven JA. Phytoplankton in a changing world: Cell size and elemental stoichiometry. J Plankton Res 2010;32:119-137.

[2] Kriest I, Oschlies A. On the treatment of particulate organic matter sinking in large-scale models of marine biogeochemical cycles. Biogeosciences 2008;5:55-72.

[3] Michaels AF, Silver MW. Primary production, sinking fluxes and the microbial food web. Deep-Sea Res I: Oceanogr Res Pap 1988;35:473-490.

[4] Giering SLC, Cavan EL, Basedow SL, Briggs N, Burd AB, Darroch LJ, et al. Sinking organic particles in the ocean—flux estimates from in situ optical devices. Front Mar Sci 2020;6:834.

[5] Huisman J, Sommeijer B. Population dynamics of sinking phytoplankton in light-limited environments: Simulation techniques and critical parameters. J Sea Res 2002;48:83-96.

[6] Turner JT. Zooplankton fecal pellets, marine snow and sinking phytoplankton blooms. Aquat Microb Ecol 2002;27:57-102.

[7] Matek A, Bosak S, Šupraha L, Neeley A, Višić H, Cetinić I, et al. Phytoplankton diversity and chemotaxonomy in contrasting North Pacific ecosystems. PeerJ 2023;11:e14501.

[8] Li Z, Sun D, Wang S, Huan Y, Zhang H, Liu J, et al. A global satellite observation of phytoplankton taxonomic groups over the past two decades. Glob Chang Biol 2023;29:4511–4529.

[9] Rousseaux CS, Gregg WW. Interannual variation in phytoplankton primary production at a global scale. Remote Sens (Basel) 2013;6:1–19.
[10] Kilham P, Hecky RE. Comparative ecology of marine and freshwater phytoplankton. Limnol Oceanogr 1988;33:776–795.
[11] Raven JA, Beardall J. Evolution of phytoplankton in relation to their physiological traits. J Mar Sci Eng 2022;10:194.
[12] Walsby AE. Gas vesicles. Microbiol Rev 1994;58:94-144.
[13] Purcell EM. Life at low Reynolds number. Am J Phys 1977;45:3-11.
[14] Hitchcock GL. A comparative study of the size-dependent organic composition of marine diatoms and dinoflagellates. J Plankton Res 1982;4:363-377.
[15] Tang EPY. Why do dinoflagellates have lower growth rates? J Phycol 1996;32:80-84.
[16] Ross ON, Sharples J. Phytoplankton motility and the competition for nutrients in the thermocline. Mar Ecol Prog Ser 2007;347:21-38.
[17] Nasouri B, Vilfan A, Golestanian R. Minimum dissipation theorem for microswimmers. Phys Rev Lett 2021;126:034503.
[18] Padisák J, Soróczki-Pintér É, Rezner Z. Sinking properties of some phytoplankton shapes and the relation of form resistance to morphological diversity of plankton - An experimental study. Hydrobiologia 2003;500:243-257.
[19] Purcell EM. The efficiency of propulsion by a rotating flagellum. Proc Natl Acad Sci USA 1997;94: 11307-11311.
[20] Tam D, Hosoi AE. Optimal feeding and swimming gaits of biflagellated organisms. Proc Natl Acad Sci USA 2011;108:1001-1006.
[21] Barsanti L, Coltelli P, Evangelista V, Frassanito AM, Gualtieri P. Swimming patterns of the quadriflagellate *Tetraflagellochloris mauritanica* (Chlamydomonadales, Chlorophyceae). J Phycol 2016;52:209-218.
[22] Chen DTN, Heymann M, Fraden S, Nicastro D, Dogic Z. ATP consumption of eukaryotic flagella measured at a single-cell level. Biophys J 2015;109:2562-2573.
[23] Spagnolie SE, Lauga E. The optimal elastic flagellum. Phys Fluid 2010;22:031901.
[24] Schavemaker PE, Lynch M. Flagellar energy costs across the tree of life. eLife 2022;11:e77266.
[25] Raven JA, Lavoie M. Movement of aquatic oxygenic photosynthetic organisms, In: Progress in botany, Springer, 2021;83:315-343.
[26] Miklasz KA, Denny MW. Diatom sinking speeds: Improved predictions and insight from a modified Stoke's law. Limnol Oceanogr 2010;55:2513-2525.
[27] Raven JA, Waite AM. The evolution of silicification in diatoms: Inescapable sinking and sinking as escape? New Phytol 2004;162:45-61.
[28] Finkel Z V., Kotrc B. Silica use through time: Macroevolutionary change in the morphology of the diatom frustule. Geomicrobiol J 2010;27:596-608.
[29] Lisicki M, Velho Rodrigues MF, Goldstein RE, Lauga E. Swimming eukaryotic microorganisms exhibit a universal speed distribution. eLife 2019;8:e44907.
[30] Geyer VF, Jülicher F, Howard J, Friedrich BM. Cell-body rocking is a dominant mechanism for flagellar synchronization in a swimming alga. Proc Natl Acad Sci USA 2013;110:18058-18063.
[31] Nielsen LT, Kiørboe T. Foraging trade-offs, flagellar arrangements, and flow architecture of planktonic protists. Proc Natl Acad Sci USA 2021;118:e2009930118.
[32] Fenchel T. How dinoflagellates swim. Protist 2001;152:329-338.
[33] Ehlers K, Oster G. On the mysterious propulsion of *Synechococcus*. PLoS One 2012;7:e36081.
[34] Stocker R, Seymour JR. Ecology and physics of bacterial chemotaxis in the ocean. Microbiol Mol Biol Rev 2012;76:792–812.
[35] Stone HA, Samuel ADT. Propulsion of microorganisms by surface distortions. Phys Rev Lett 1996;77:4102.
[36] Ehlers KM, Samuel ADT, Berg HC, Montgomery R. Do cyanobacteria swim using traveling surface waves? Proc Natl Acad Sci USA 1996;93:8340-8343.
[37] Samadi Z, Mehdizadeh Allaf M, Vourc'h T, DeGroot CT, Peerhossaini H. Investigation of *Synechocystis* sp. CPCC 534 motility during different stages of the growth period in active fluids. Processes 2023;11:1492.
[38] Maier B, Potter L, So M, Seifert HS, Sheetz MP. Single pilus motor forces exceed 100 pN. Proc Natl Acad Sci USA 2002;99:16012-16017.
[39] Clausen M, Koomey M, Maier B. Dynamics of type IV pili is controlled by switching between multiple states. Biophys J 2009;96:1169-1177.

[40] Craig L, Forest KT, Maier B. Type IV pili: Dynamics, biophysics and functional consequences. Nat Rev Microbiol 2019;17:429-440.
[41] McBride MJ. Bacterial gliding motility: Multiple mechanisms for cell movement over surfaces. Annu Rev Microbiol 2001;55:49-75.
[42] Balagam R, Litwin DB, Czerwinski F, Sun M, Kaplan HB, Shaevitz JW, et al. *Myxococcus xanthus* gliding motors are elastically coupled to the substrate as predicted by the focal adhesion model of gliding motility. PLoS Comput Biol 2014;10:e1003619.
[43] Jakobsen HH. Escape of protists in predator-generated feeding currents. Aquat Microb Ecol 2002;26:271–81.
[44] Jiang H, Kulis DM, Brosnahan ML, Anderson DM. Behavioral and mechanistic characteristics of the predator-prey interaction between the dinoflagellate *Dinophysis acuminata* and the ciliate *Mesodinium rubrum*. Harmful Algae 2018;77:43–54.
[45] Miano F, Asadzadeh SS, Ryderheim F, Andersen A, Kiørboe T. High-speed escape jumps in haptophytes: Mechanism and triggering fluid signal. Limnol Oceanogr 2024;69:2846–2858.
[46] Jiang H, Johnson MD. Jumping and overcoming diffusion limitation of nutrient uptake in the photosynthetic ciliate *Mesodinium rubrum*. Limnol Oceanogr 2017;62:421–36.
[47] Guasto JS, Rusconi R, Stocker R. Fluid mechanics of planktonic microorganisms. Annu Rev Fluid Mech 2011;44:373-400.
[48] Kundu P, Cohen I, Dowling D. Fluid Mechanics, 6th Ed., Elsevier 2012.
[49] Lynch MR. Evolutionary cell biology: The origins of cellular architecture. Oxford University Press; 2024.
[50] Lighthill MJ. On the squirming motion of nearly spherical deformable bodies through liquids at very small Reynolds numbers. Commun Pure Appl Math 1952;5:109–118.
[51] Childress S. Mechanics of swimming and flying. Cambridge University Press; 1981.
[52] Leal LG. Advanced transport phenomena: Fluid mechanics and convective transport processes. Cambridge university press; 2007.
[53] Guadayol Ò, Mendonca T, Segura-Noguera M, Wright AJ, Tassieri M, Humphries S. Microrheology reveals microscale viscosity gradients in planktonic systems. Proc Natl Acad Sci USA 2021;118:e2011389118.
[54] Stehnach MR, Waisbord N, Walkama DM, Guasto JS. Viscophobic turning dictates microalgae transport in viscosity gradients. Nat Phys 2021;17:926-930.
[55] Pozrikidis C. Boundary integral and singularity methods for linearized viscous flow. Cambridge University Press, 1992.
[56] Kiørboe T, Jiang H, Gonçalves RJ, Nielsen LT, Wadhwa N. Flow disturbances generated by feeding and swimming zooplankton. Proc Natl Acad Sci USA 2014;111:11738-11743.
[57] Andersen A, Wadhwa N, Kiørboe T. Quiet swimming at low Reynolds number. Phys Rev E 2015;91: 042712.
[58] Lauga E, Powers TR. The hydrodynamics of swimming microorganisms. Rep Prog Phys 2009;72:096601.
[59] Reynolds CS. The ecology of phytoplankton. Cambridge University Press, 2006.
[60] Moore JK, Villareal TA. Buoyancy and growth characteristics of three positively buoyant marine diatoms. Mar Ecol Prog Ser 1996;132:203-213.
[61] Woods S, Villareal TA. Intracellular ion concentrations and cell sap density in positively buoyant oceanic phytoplankton. Nova Hedwigia 2008;133:131-145.
[62] Boyd CM, Gradmann D. Impact of osmolytes on buoyancy of marine phytoplankton. Mar Biol 2002;141:605-618.
[63] Wirtz K, Smith SL. Vertical migration by bulk phytoplankton sustains biodiversity and nutrient input to the surface ocean. Sci Rep 2020;10:1142.
[64] Kromkamp JC, Mur LR. Buoyant density changes in the cyanobacterium *Microcystis aeruginosa* due to changes in the cellular carbohydrate content. FEMS Microbiol Lett 1984;25:105–109.
[65] Feng G, Visser PM, Huisman J, Verspagen JMH. Rising temperature accelerates buoyancy regulation and vertical migration of the bloom-forming cyanobacterium *Microcystis*. Water Res 2025;286:124259.
[66] Yoshiyama K, Sawa T, Yoshiyama Y. Fundamental unpredictability in the vertical migration of cyanobacteria. Theor Ecol 2025;18:12.
[67] Lavoie M, Raven JA, Levasseur M. Energy cost and putative benefits of cellular mechanisms modulating buoyancy in aflagellate marine phytoplankton. J Phycol 2016;52:239-251.

[68] Raven JA, Doblin MA. Active water transport in unicellular algae: Where, why, and how. J Exp Bot 2014;65:6279-6292.
[69] Petrucciani A, Moretti P, Ortore MG, Norici A. Integrative effects of morphology, silicification, and light on diatom vertical movements. Front Plant Sci 2023;14:1143998.
[70] Richardson TL, Cullen JJ. Changes in buoyancy and chemical composition during growth of a coastal marine diatom: Ecological and biogeochemical consequences. Mar Ecol Prog Ser 1995;128:77–90.
[71] Richardson TL, Ciotti ÁM, Cullen JJ, Villareal TA. Physiological and optical properties of *Rhizosolenia formosa* (bacillariophyceae) in the context of open-ocean vertical migration. J Phycol 1996;32:741–757.
[72] Moore JK, Villareal TA. Buoyancy and growth characteristics of three positively buoyant marine diatoms. Mar Ecol Prog Ser 1996;132:203–213.
[73] Acuña JL, López-Alvarez M, Nogueira E, González-Taboada F. Diatom flotation at the onset of the spring phytoplankton bloom: An in situ experiment. Mar Ecol Prog Ser 2010;400:115–125.
[74] Durante G, Basset A, Stanca E, Roselli L. Allometric scaling and morphological variation in sinking rate of phytoplankton. J Phycol 2019;55:1386-1393.
[75] Gemmell BJ, Oh G, Buskey EJ, Villareal TA. Dynamic sinking behaviour in marine phytoplankton: Rapid changes in buoyancy may aid in nutrient uptake. Proc Roy Soc B 2016;283:20161126.
[76] Du Clos KT, Karp-Boss L, Gemmell BJ. Diatoms rapidly alter sinking behavior in response to changing nutrient concentrations. Limnol Oceanogr 2021;66:892-900.
[77] Du Clos KT, Karp-Boss L, Villareal TA, Gemmell BJ. *Coscinodiscus wailesii* mutes unsteady sinking in dark conditions. Biol Lett 2019;15:20180816.
[78] Behrenfeld MJ, Halsey KH, Boss E, Karp-Boss L, Milligan AJ, Peers G. Thoughts on the evolution and ecological niche of diatoms. Ecol Monogr 2021;91:e01457.
[79] Raven JA. Scaling the seas. Plant Cell Environ 1995;18:1090-1100.
[80] Zhang W, Hao Q, Zhu J, Deng Y, Xi M, Cai Y, et al. Nanoplanktonic diatom rapidly alters sinking velocity via regulating lipid content and composition in response to changing nutrient concentrations. Front Mar Sci 2023;10:1255915.
[81] Wu Y, Kouadio VK, Usherwood TR, Li J, Bisher M, Aurora R, et al. Diverse biophysical and molecular mechanisms drive phytoplankton sinking in response to starvation. PLoS Biol 2025;23:e3003508.
[82] Raven JA. The role of vacuoles. New Phytol 1987;106:357-422.
[83] Lavoie M, Raven JA. How can large-celled diatoms rapidly modulate sinking rates episodically? J Exp Bot 2020;71:3386-3389.
[84] Larson AG, Chajwa R, Li H, Prakash M. Inflation-induced motility for long-distance vertical migration. Curr Biol 2024;34:5149-5163.
[85] Naselli-Flores L, Barone R. Fight on plankton! Or, phytoplankton shape and size as adaptive tools to get ahead in the struggle for life. Cryptogam Algol 2011;32:157-204.
[86] Raven JA, Richardson K. Dinophyte flagella: A cost-benefit analysis. New Phytol 1984;98:259-276.
[87] Ji R, Franks PJS. Vertical migration of dinoflagellates: Model analysis of strategies, growth, and vertical distribution patterns. Mar Ecol Prog Ser 2007;344:49-61.
[88] Cullen JJ, Horrigan SG. Effects of nitrate on the diurnal vertical migration, carbon to nitrogen ratio, and the photosynthetic capacity of the dinoflagellate *Gymnodinium splendens*. Mar Biol 1981;62:81–89.
[89] Durham WM, Stocker R. Thin phytoplankton layers: Characteristics, mechanisms, and consequences. Ann Rev Mar Sci 2012;4:177–207.
[90] Cullen JJ, MacIntyre JG. Behavior, physiology and the niche of depth-regulating phytoplankton. Nato Asi Ser G Ecol Sci 1998;41:559–580.
[91] Schuech R, Menden-Deuer S. Going ballistic in the plankton: Anisotropic swimming behavior of marine protists. Limnol Oceanogr: Fluids Environ 2014;4:1-16.
[92] Sala I, Vallina SM, Lévy M, Bolado-Penagos M, García CM, Echevarría F, et al. Modelling the effect of the tidal cycle on the high phytoplankton biomass area of Cape Trafalgar (SW Iberian Peninsula). Prog Oceanogr 2023;217:103085.
[93] Wirtz K, Smith SL, Mathis M, Taucher J. Vertically migrating phytoplankton fuel high oceanic primary production. Nat Clim Chang 2022;12:750-756.
[94] Cullen JJ. Subsurface chlorophyll maximum layers: Enduring enigma or mystery solved? Ann Rev Mar Sci 2015;7:207-239.

[95] Klausmeier CA, Litchman E. Algal games: The vertical distribution of phytoplankton in poorly mixed water columns. Limnol Oceanogr 2001;46:1998-2007.
[96] Lévy M, Jahn O, Dutkiewicz S, Follows MJ. Phytoplankton diversity and community structure affected by oceanic dispersal and mesoscale turbulence. Limnol Oceanogr: Fluids Environ 2014;4:67-84.
[97] Aghamohammadi A, Aghamohammadi C, Moghimi-Araghi S. On swimmer's strategies in various currents. Eur J Phys 2023;44:055002.
[98] Smayda TJ. Adaptations and selection of harmful and other dinoflagellate species in upwelling systems. 2. Motility and migratory behaviour. Prog Oceanogr 2010;85:71-91.
[99] Sullivan JM, Swift E, Donaghay PL, Rines JEB. Small-scale turbulence affects the division rate and morphology of two red-tide dinoflagellates. Harmful Algae 2003;2:183-199.
[100] Lovecchio S, Climent E, Stocker R, Durham WM. Chain formation can enhance the vertical migration of phytoplankton through turbulence. Sci Adv 2019;5:eaaw7879.
[101] Fraga S, Gallager SM, Anderson DM. Chain-forming dinoflagellates: An adaptation to red tides. In: Red Tides: Biological, Environmental Science, and Toxicology, 1989;281-284.
[102] Sengupta A, Carrara F, Stocker R. Phytoplankton can actively diversify their migration strategy in response to turbulent cues. Nature 2017;543:555-558.
[103] Jékely G. Evolution of phototaxis. Phil Transac Roy Soc B 2009;364:2795–2808.
[104] Blackburn N, Fenchel T. Modelling of microscale patch encounter by chemotactic protozoa. Protist 1999;150:337–343.
[105] Govorunova EG, Sineshchekov OA. Chemotaxis in the green flagellate alga *Chlamydomonas*. Biochemistry (Moscow) 2005;70:717–725.
[106] Langlois VJ, Andersen A, Bohr T, Visser AW, Kiørboe T. Significance of swimming and feeding currents for nutrient uptake in osmotrophic and interception-feeding flagellates. Aquat Microb Ecol 2009;54:35-44.
[107] Nielsen LT, Kiørboe T. Feeding currents facilitate a mixotrophic way of life. ISME J 2015;9:2117-2127.
[108] Magar V, Goto T, Pedley TJ. Nutrient uptake by a self-propelled steady squirmer. Qtr J Mech Appl Math 2003;56:65-91.
[109] Clift R, Grace JR, Weber ME. Bubbles, drops and particles. Academic Press 1978;10.
[110] Princiotta MF, Finzi D, Qian S-B, Gibbs J, Schuchmann S, Buttgereit F, et al. Quantitating protein synthesis, degradation, and endogenous antigen processing. Immunity 2003;18:343–354.
[111] Nascimento JF, Souza ROO, Alencar MB, Marsiccobetre S, Murillo AM, Damasceno FS, et al. How much (ATP) does it cost to build a trypanosome? A theoretical study on the quantity of ATP needed to maintain and duplicate a bloodstream-form *Trypanosoma brucei* cell. PLoS Pathog 2023;19:e1011522.
[112] Milo R, Phillips R. Cell biology by the numbers. 1st Ed. Taylor and Francis 2015.
[113] Zhu S, Gao B. Bacterial flagella loss under starvation. Trends Microbiol 2020;28:785-788.
[114] Ferreira JL, Gao FZ, Rossmann FM, Nans A, Brenzinger S, Hosseini R, et al. γ-proteobacteria eject their polar flagella under nutrient depletion, retaining flagellar motor relic structures. PLoS Biol 2019;17: e3000165.
[115] Bauer D, Ishikawa H, Wemmer KA, Hendel NL, Kondev J, Marshall WF. Analysis of biological noise in the flagellar length control system. iScience 2021;24:102354.
[116] Nguyen RL, Tam LW, Lefebvre PA. The LF1 gene of *Chlamydomonas reinhardtii* encodes a novel protein required for flagellar length control. Genetics 2005;169:1415-1424.
[117] Samuel ADT, Petersen JD, Reese TS. Envelope structure of *Synechococcus* sp. WH8113, a nonflagellated swimming cyanobacterium. BMC Microbiol 2001;1:4.
[118] Brahamsha B. An abundant cell-surface polypeptide is required for swimming by the nonflagellated marine cyanobacterium *Synechococcus*. Proc Natl Acad Sci USA 1996;93:6504-6509.
[119] McCarren J, Brahamsha B. SwmB, a 1.12-megadalton protein that is required for nonflagellar swimming motility in *Synechococcus*. J Bacteriol 2007;189:1158-1162.
[120] Chang YW, Rettberg LA, Treuner-Lange A, Iwasa J, Søgaard-Andersen L, Jensen GJ. Architecture of the type IVa pilus machine. Science 2016;351:468a-469a.
[121] Chandra A, Joubert L-M, Bhaya D. Modulation of type IV pili phenotypic plasticity through a novel Chaperone-Usher system in *Synechocystis* sp. BioRxiv 2017:1–17.

[122] Monteiro FM, Bach LT, Brownlee C, Bown P, Rickaby REM, Poulton AJ, et al. Why marine phytoplankton calcify. Sci Adv 2016;2:e1501822.
[123] Stone HA, Samuel ADT. Propulsion of microorganisms by surface distortions. Phys Rev Lett 1996;77:4102.
[124] Chattopadhyay S, Moldovan R, Yeung C, Wu XL. Swimming efficiency of bacterium *Escherichia coli*. Proc Natl Acad Sci USA 2006;103:13712-13717.
[125] Arroyo M, Heltai L, Millán D, DeSimone A. Reverse engineering the euglenoid movement. Proc Natl Acad Sci USA 2012;109:17874-17879.
[126] Piro L, Vilfan A, Golestanian R, Mahault B. Energetic cost of microswimmer navigation: The role of body shape. Phys Rev Res 2024;6: 013274.
[127] Guo H, Zhu H, Liu R, Bonnet M, Veerapaneni S. Optimal slip velocities of micro-swimmers with arbitrary axisymmetric shapes. J Fluid Mech 2021;910:A26.
[128] Vilfan A. Optimal shapes of surface slip driven self-propelled microswimmers. Phys Rev Lett 2012;109:128105.
[129] Happel J, Brenner H. Low Reynolds number hydrodynamics: with special applications to particulate media. Springer Science & Business Media; 2012.
[130] Giri P, Shukla RK. Optimal transport of surface-actuated microswimmers. Phys Fluids 2022;34:043604.
[131] Ryabov A, Kerimoglu O, Litchman E, Olenina I, Roselli L, Basset A, et al. Shape matters: The relationship between cell geometry and diversity in phytoplankton. Ecol Lett 2021;24:847-861.
[132] Jeffery GB. The motion of ellipsoidal particles immersed in a viscous fluid. Proc Roy Soc A, 1922;102:161–79.
[133] Šmarda J, Šmajs D, Komrska J, Krzyžánek V. S-layers on cell walls of cyanobacteria. Micron 2002;33:257-277.
[134] Jastroch M, Divakaruni AS, Mookerjee S, Treberg JR, Brand MD. Mitochondrial proton and electron leaks. Essays Biochem 2010;47:53–67.
[135] Bertholet AM, Natale AM, Bisignano P, Suzuki J, Fedorenko A, Hamilton J, et al. Mitochondrial uncouplers induce proton leak by activating AAC and UCP1. Nature 2022;606:180–187.
[136] Divakaruni AS, Brand MD. The regulation and physiology of mitochondrial proton leak. Physiology (Bethesda) 2011;26:192–205.
[137] Ng D, Harn T, Altindal T, Kolappan S, Marles JM, Lala R, et al. The vibrio cholerae minor pilin TcpB initiates assembly and retraction of the toxin-coregulated pilus. PLoS Pathog 2016;12: e1006109.
[138] Jiang H. Why does the jumping ciliate *Mesodinium rubrum* possess an equatorially located propulsive ciliary belt? J Plankton Res 2011;33:998–1011.
[139] Naselli-Flores L, Zohary T, Padisák J. Life in suspension and its impact on phytoplankton morphology: An homage to Colin S. Reynolds. Hydrobiologia 2021;848:7-30.
[140] O'Farrell I, Pinto PDT, Izaguirre I. Phytoplankton morphological response to the underwater light conditions in a vegetated wetland. Hydrobiologia 2007;578:65-77.
[141] Kléparski L, Beaugrand G, Edwards M, Schmitt FG, Kirby RR, Breton E, et al. Morphological traits, niche-environment interaction and temporal changes in diatoms. Prog Oceanogr 2022;201:102747.
[142] Huber ST, Terwiel D, Evers WH, Maresca D, Jakobi AJ. Cryo-EM structure of gas vesicles for buoyancy-controlled motility. Cell 2023;186:975-986.
[143] Pfeifer F. Distribution, formation and regulation of gas vesicles. Nat Rev Microbiol 2012;10:705-715.
[144] Walsby AE, Bleything A. The dimensions of cyanobacterial gas vesicles in relation to their efficiency in providing buoyancy and withstanding pressure. Microbiology 1988;134: 2635-2645.
[145] Bhattacharya R, Lin SGM, Basu NB. Windows into the past: lake sediment phosphorus trajectories act as integrated archives of watershed disturbance legacies over centennial scales. Environ Res Lett 2022;17:034005.
[146] Abeynayaka HDL, Asaeda T, Kaneko Y. Buoyancy limitation of filamentous cyanobacteria under prolonged pressure due to the gas vesicles collapse. Environ Manage 2017;60: 293-303.
[147] Walsby AE. The mechanical properties of the *Microcystis* gas vesicle. J Gen Microbiol 1991;137:2401-2408.
[148] Hoffmann R, Kirchlechner C, Langer G, Wochnik AS, Griesshaber E, Schmahl WW, et al. Insight into *Emiliania huxleyi* coccospheres by focused ion beam sectioning. Biogeosciences 2015;12:825-834.
[149] Rost B, Riebesell U. Coccolithophores and the biological pump: Responses to environmental changes. In: Coccolithophores, Springer 2004;99-125.

[150] Young JR. Possible functional interpretations of coccolith morphology. Abh Geol B-A 1987;39:305-313.
[151] Paasche E. A review of the coccolithophorid *Emiliania huxleyi* (prymnesiophyceae), with particular reference to growth, coccolith formation, and calcification-photosynthesis interactions. Phycologia 2001;40:503-529.
[152] Jaya BN, Hoffmann R, Kirchlechner C, Dehm G, Scheu C, Langer G. Coccospheres confer mechanical protection: New evidence for an old hypothesis. Acta Biomater 2016;42:258-264.
[153] Amirian MM, Finkel Z V, Devred E, Irwin AJ. Parameterization of photoinhibition for phytoplankton. Commun Earth Environ 2025;6:707.
[154] Katsu-Kimura Y, Nakaya F, Baba SA, Mogami Y. Substantial energy expenditure for locomotion in ciliates verified by means of simultaneous measurement of oxygen consumption rate and swimming speed. J Exp Biol 2009;212:1819–1824.
[155] Mookerjee SA, Gerencser AA, Nicholls DG, Brand MD. Quantifying intracellular rates of glycolytic and oxidative ATP production and consumption using extracellular flux measurements. J Biol Chem 2017;292: 7189-7207.
[156] van Tol HM, Irwin AJ, Finkel Z V. Macroevolutionary trends in silicoflagellate skeletal morphology: The costs and benefits of silicification. Paleobiology 2012;38:391-402.
[157] Petrou K, Baker KG, Nielsen DA, Hancock AM, Schulz KG, Davidson AT. Acidification diminishes diatom silica production in the Southern Ocean. Nat Clim Chang 2019;9:781-786.
[158] Bidle KD, Azam F. Accelerated dissolution of diatom silica by marine bacterial assemblages. Nature 1999;397:508-512.
[159] Patrick S, Holding AJ. The effect of bacteria on the solubilization of silica in diatom frustules. J Appl Bacteriol 1985;59:7-16.
[160] Raven JA. The energetics of freshwater algae; energy requirements for biosynthesis and volume regulation. New Phytol 1982;92:1–20.
[161] Friedrichs L, Hörnig M, Schulze L, Bertram A, Jansen S, Hamm C. Size and biomechanic properties of diatom frustules influence food uptake by copepods. Mar Ecol Prog Ser 2013;481:41-51.
[162] Blaber S, Louwerse MD, Sivak DA. Steps minimize dissipation in rapidly driven stochastic systems. Phys Rev E 2021;104: L022101.
[163] Ješková Z, Featonby D, Feková V. Balloons revisited. Phys Educ 2012;47:392.
[164] Vandermarlière J. On the inflation of a rubber balloon. Phys Teach 2016;54:566-567.
[165] Stein RS. On the inflating of balloons. J Chem Educ 1958;35:203.
[166] Meunier CL, Schulz K, Boersma M, Malzahn AM. Impact of swimming behaviour and nutrient limitation on predator–prey interactions in pelagic microbial food webs. J Exp Mar Biol Ecol 2013;446:29–35.
[167] Visser AW. Motility of zooplankton: fitness, foraging and predation. J Plankton Res 2007;29:447–461.
[168] Visser AW, Mariani P, Pigolotti S. Swimming in turbulence: zooplankton fitness in terms of foraging efficiency and predation risk. J Plankton Res 2009;31:121–133.
[169] Kiørboe T. Predation in a microbial world: Mechanisms and trade-offs of flagellate foraging. Ann Rev Mar Sci 2024;16:361-381.
[170] Chakraborty S, Pančić M, Andersen KH, Kiørboe T. The cost of toxin production in phytoplankton: The case of PST producing dinoflagellates. ISME J 2019;13:64-75.
[171] Haddock SHD, Moline MA, Case JF. Bioluminescence in the sea. Ann Rev Mar Sci 2010;2:443-493.
[172] Valiadi M, Iglesias-Rodriguez D. Understanding bioluminescence in dinoflagellates—how far have we come? Microorganisms 2013;1:3-25.
[173] Cusick KD, Widder EA. Bioluminescence and toxicity as driving factors in harmful algal blooms: Ecological functions and genetic variability. Harmful Algae 2020;98:101850.
[174] Hastings JW. Bacterial and dinoflagellate luminescent systems. In: Bioluminescence in Action. Herring PJ, Academic Press 1978.
[175] Marcinko CLJ, Painter SC, Martin AP, Allen JT. A review of the measurement and modelling of dinoflagellate bioluminescence. Prog Oceanogr 2013;109:117-129.
[176] Levin BR, Stewart FM, Chao L. Resource-limited growth, competition, and predation: a model and experimental studies with bacteria and bacteriophage. Am Nat 1977;111:3–24.
[177] Bohannan BJM, Lenski RE. Linking genetic change to community evolution: insights from studies of bacteria and bacteriophage. Ecol Lett 2000;3:362–377.

[178] Lawrence JE, Suttle CA. Effect of viral infection on sinking rates of *Heterosigma akashiwo* and its implications for bloom termination. Aquat Microb Ecol 2004;37:1-7.
[179] Truchon AR, Chase EE, Stark AR, Wilhelm SW. The diel disconnect between cell growth and division in *Aureococcus* is interrupted by giant virus infection. Front Microbiol 2024;15:1426193.
[180] Danovaro R, Corinaldesi C, Dell'Anno A, Fuhrman JA, Middelburg JJ, Noble RT, et al. Marine viruses and global climate change. FEMS Microbiol Rev 2011;35:993-1034.
[181] Lohrmann C, Holm C, Datta SS. Influence of bacterial swimming and hydrodynamics on infection by phages. BioRxiv 2024. https://doi.org/10.1101/2024.01.15.575727
[182] Olive M, Moerman F, Fernandez-Cassi X, Altermatt F, Kohn T. Removal of waterborne viruses by tetrahymena pyriformis is virus-specific and coincides with changes in protist swimming speed. Environ Sci Technol 2022;56:4062-4070.
[183] Fenchel T, Finlay BJ. Respiration rates in heterotrophic, free-living protozoa. Microb Ecol 1983;9:99-122.
[184] Crawford DW. Metabolic cost of motility in planktonic protists: Theoretical considerations on size scaling and swimming speed. Microb Ecol 1992:1–10.
[185] DeLong JP, Okie JG, Moses ME, Sibly RM, Brown JH. Shifts in metabolic scaling, production, and efficiency across major evolutionary transitions of life. Proc Natl Acad Sci USA 2010;107: 12941-12945.
[186] Fahimi P, Matta CF, Okie JG. Are size and mitochondrial power of cells inter-determined? J Theor Biol 2023;572:111565.
[187] Fahimi P, Matta CF. On the power per mitochondrion and the number of associated active ATP synthases. Phys Biol 2021;18: 04LT01.
[188] Kliphuis AMJ, Janssen M, van den End EJ, Martens DE, Wijffels RH. Light respiration in *Chlorella sorokiniana*. J Appl Phycol 2011;23:935-947.
[189] Nicholls DG, Ferguson SJ. Bioenergetics. 4th Ed. Academic Press 2013.
[190] Schavemaker PE, Lynch M. Bioenergetics and the evolution of cellular traits. Annu Rev Biophys 2025;54:81–99.
[191] Tomas CR. Identifying marine phytoplankton. Elsevier; 1997.
[192] Heil CA, Glibert PM, Fan C. *Prorocentrum minimum* (Pavillard) Schiller: A review of a harmful algal bloom species of growing worldwide importance. Harmful Algae 2005;4:449-470.
[193] Kamykowski D, Reed RE, Kirkpatrick GJ. Comparison of sinking velocity, swimming velocity, rotation and path characteristics among six marine dinoflagellate species. Mar Biol 1992;113:319-328.
[194] Finkel Z V., Follows MJ, Irwin AJ. Size-scaling of macromolecules and chemical energy content in the eukaryotic microalgae. J Plankton Res 2016;38:1151-1162.
[195] Tang EPY. The allometry of algal growth and respiration. MSc Thesis, McGill University 1995.
[196] Brand MD, Brindle KM, Buckingham JA, Harper JA, Rolfe DFS, Stuart JA. The significance and mechanism of mitochondrial proton conductance. Int J Obes 1999;23:S4–S11.
[197] Brand MD. The efficiency and plasticity of mitochondrial energy transduction. Biochem Soc Trans 2005;33: 897–904.
[198] Geider RJ, Osborne BA. Respiration and microalgal growth: A review of the quantitative relationship between dark respiration and growth. New Phytol 1989;112:327-341.
[199] Langdon C. The significance of respiration in production measurements based on oxygen. ICES Mar. Sci. Symp, vol. 197, 1993, p. 69–78.
[200] Richter M, Daufenbach J, Drebing S, Vucetic V, Nguyen DT. Light-induced proton slip and proton leak at the thylakoid membrane. J Plant Physiol 2004;161: 1325-1337.
[201] Buchert F, Bailleul B, Joliot P. Disentangling chloroplast ATP synthase regulation by proton motive force and thiol modulation in *Arabidopsis* leaves. Biochim Biophys Acta Bioenerg 2021;1862: 148434.
[202] Allen JF. Photosynthesis of ATP-electrons, proton pumps, rotors, and poise. Cell 2002;110:273-276.
[203] Burlacot A. Quantifying the roles of algal photosynthetic electron pathways: A milestone towards photosynthetic robustness. New Phytol 2023;240: 2197-2203.
[204] Fahimi P, Amirian MM, Irwin AJ, Finkel Z V. Carbon-to-ATP ratios across the kingdoms of life. iScience 2026;29:116849.
[205] Bailleul B, Berne N, Murik O, Petroutsos D, Prihoda J, Tanaka A, et al. Energetic coupling between plastids and mitochondria drives $CO_2$ assimilation in diatoms. Nature 2015;524:366-369.

[206] López-Sandoval DC, Rodríguez-Ramos T, Cermeño P, Sobrino C, Marañón E. Photosynthesis and respiration in marine phytoplankton: Relationship with cell size, taxonomic affiliation, and growth phase. J Exp Mar Biol Ecol 2014;457:151-159.
[207] Hansen PJ, Bjørnsen PK, Hansen BW. Zooplankton grazing and growth: Scaling within the 2-2,000-µm body size range. Limnol Oceanogr 1997;42:687–704.
[208] Jeong HJ, Du Yoo Y, Kim ST, Kang NS. Feeding by the heterotrophic dinoflagellate *Protoperidinium bipes* on the diatom *Skeletonema costatum*. Aquat Microb Ecol 2004;36:171–179.
[209] Baek SH, Shimode S, Shin K, Han M-S, Kikuchi T. Growth of dinoflagellates, *Ceratium furca* and *Ceratium fusus* in Sagami Bay, Japan: The role of vertical migration and cell division. Harmful Algae 2009;8:843–856.
[210] Burov ON, Kletskii ME, Kurbatov S V., Lisovin A V., Fedik NS. Mechanisms of nitric oxide generation in living systems. Nitric Oxide – Biol Chem 2022;118:1-16.
[211] Doney SC. Oceanography: Plankton in a warmer world. Nature 2006;444:695-696.
[212] Regaudie-De-Gioux A, Duarte CM. Temperature dependence of planktonic metabolism in the ocean. Glob Biogeochem Cycl 2012;26: GB1015.
[213] Sekerci Y, Petrovskii S. Mathematical modelling of plankton–oxygen dynamics under the climate change. Bull Math Biol 2015;77: 2325-2353.
[214] Sheward RM, Liefer JD, Irwin AJ, Finkel Z V. Elemental stoichiometry of the key calcifying marine phytoplankton *Emiliania huxleyi* under ocean climate change: A meta-analysis. Glob Chang Biol 2023;29: 4259-4278.
[215] McGinty N, Barton AD, Record NR, Finkel Z V., Johns DG, Stock CA, et al. Anthropogenic climate change impacts on copepod trait biogeography. Glob Chang Biol 2021;27: 1431-1442.
[216] Barton AD, Irwin AJ, Finkel Z V., Stock CA. Anthropogenic climate change drives shift and shuffle in North Atlantic phytoplankton communities. Proc Natl Acad Sci USA 2016;113: 2964-2969.
[217] Lotze HK, Mellon S, Coyne J, Betts M, Burchell M, Fennel K, et al. Long-term ocean and resource dynamics in a hotspot of climate change. Facets 2022;7: 1142-1184.
[218] Heneghan RF, Galbraith E, Blanchard JL, Harrison C, Barrier N, Bulman C, et al. Disentangling diverse responses to climate change among global marine ecosystem models. Prog Oceanogr 2021;198: 102659.
[219] Mutshinda CM, Finkel ZV, Irwin AJ. Large, regionally variable shifts in diatom and dinoflagellate biomass in the North Atlantic over six decades. PLoS One 2025;20:e0323675.
[220] Svensson F, Norberg J, Snoeijs P. Diatom cell size, coloniality and motility: Trade-offs between temperature, salinity and nutrient supply with climate change. PLoS One 2014;9: e109993.
[221] Wang Y, Fan X, Gao G, Beardall J, Inaba K, Hall-Spencer JM, et al. Decreased motility of flagellated microalgae long-term acclimated to $CO_2$-induced acidified waters. Nat Clim Chang 2020;10:561-567.
[222] Escobar A, Negro V, López-Gutiérrez JS, Esteban MD. Influence of temperature and salinity on hydrodynamic forces. J Ocean Eng Sci 2016;1:325-336.
[223] Zohary T, Flaim G, Sommer U. Temperature and the size of freshwater phytoplankton. Hydrobiologia 2021;848:143-155.
[224] Atkinson D, Ciotti BJ, Montagnes DJS. Protists decrease in size linearly with temperature: ca. 2.5% $°C^{-1}$. Proc Roy Soc B 2003;270: 2605-2611.
[225] Kamykowski D, Zentara SJ, Morrison JM, Switzer AC. Dynamic global patterns of nitrate, phosphate, silicate, and iron availability and phytoplankton community composition from remote sensing data. Glob Biogeochem Cycl 2002;16:1077.
[226] Marañón E, Cermeño P, Latasa M, Tadonléké RD. Temperature, resources, and phytoplankton size structure in the ocean. Limnol Oceanogr 2012;57: 1266-1278.
[227] Schaum CE, Buckling A, Smirnoff N, Studholme DJ, Yvon-Durocher G. Environmental fluctuations accelerate molecular evolution of thermal tolerance in a marine diatom. Nat Commun 2018;9:1719.
[228] Leles SG, Levine NM. Mechanistic constraints on the trade-off between photosynthesis and respiration in response to warming. Sci Adv 2023;9: eadh8043.
[229] Beveridge OS, Petchey OL, Humphries S. Mechanisms of temperature-dependent swimming: The importance of physics, physiology and body size in determining protist swimming speed. J Exp Biol 2010;213: 4223-4231.

[230] Bach LT, Riebesell U, Sett S, Febiri S, Rzepka P, Schulz KG. An approach for particle sinking velocity measurements in the 3-400 μm size range and considerations on the effect of temperature on sinking rates. Mar Biol 2012;159: 1853-1864.
[231] Kamykowski D, Mccollum SA. The temperature acclimatized swimming speed of selected marine dinoflagellates. J Plankton Res 1986;8: 275-287.
[232] Hansen PJ, Bjørnsen PK, Hansen BW. Zooplankton grazing and growth: Scaling within the 2-2,000-μm body size range. Limnol Oceanogr 1997;42: 687-704.
[233] Abele D, Heise K, Portner HO, Puntarulo S. Temperature-dependence of mitochondrial function and production of reactive oxygen species in the intertidal mud clam Mya arenaria. J Exp Biol 2002;205:1831–1841.
[234] Kerbler SM, Taylor NL, Millar AH. Cold sensitivity of mitochondrial ATP synthase restricts oxidative phosphorylation in *Arabidopsis thaliana*. New Phytol 2019;221: 1776-1788.
[235] Tan YF, Millar AH, Taylor NL. Components of mitochondrial oxidative phosphorylation vary in abundance following exposure to cold and chemical stresses. J Proteome Res 2012;11: 3860-3879.
[236] Padfield D, Yvon-Durocher G, Buckling A, Jennings S, Yvon-Durocher G. Rapid evolution of metabolic traits explains thermal adaptation in phytoplankton. Ecol Lett 2016;19:133-142.
[237] Marañón E, Lorenzo MP, Cermeño P, Mouriño-Carballido B. Nutrient limitation suppresses the temperature dependence of phytoplankton metabolic rates. ISME J 2018;12:1836–1845.
[238] Huang B, Rifkin MR, Luck DJL. Temperature sensitive mutations affecting flagellar assembly and function in *Chlamydomonas reinhardtii*. J Cell Biol 1977;72:67-85.
[239] Yadav S, Kunwar A. Temperature-dependent activity of motor proteins: Energetics and their implications for collective behavior. Front Cell Dev Biol 2021;9: 610899.
[240] Lee S, Kim DH, Needham D. Equilibrium and dynamic interfacial tension measurements at microscopic interfaces using a micropipet technique - 2. Dynamics of phospholipid monolayer formation and equilibrium tensions at the water-air interface. Langmuir 2001;17: 5544-5550.
[241] Kranenburg M, Smit B. Phase behavior of model lipid bilayers. J Phys Chem B 2005;109: 6553-6563.
[242] Waugh R, Evans EA. Thermoelasticity of red blood cell membrane. Biophys J 1979;26:115-131.
[243] Kirsch SA, Böckmann RA. Coupling of membrane nanodomain formation and enhanced electroporation near phase transition. Biophys J 2019;116: 2131-2148.
[244] Fang C, Ji F, Shu Z, Gao D. Determination of the temperature-dependent cell membrane permeabilities using microfluidics with integrated flow and temperature control. Lab Chip 2017;17: 951-960.
[245] Humphries S. A physical explanation of the temperature dependence of physiological processes mediated by cilia and flagella. Proc Natl Acad Sci USA 2013;110: 14693-14698.
[246] Belas R. Biofilms, flagella, and mechanosensing of surfaces by bacteria. Trends Microbiol 2014;22:517-527.
[247] Anderson JK, Smith TG, Hoover TR. Sense and sensibility: Flagellum-mediated gene regulation. Trends Microbiol 2010;18:30-37.
[248] Moens S, Vanderleyden J. Functions of bacterial flagella. Crit Rev Microbiol 1996;22:67-100.
[249] Pozrikidis C. Boundary integral and singularity methods for linearized viscous flow. Cambridge University Press 1992.
[250] Lynch M, Trickovic B. A theoretical framework for evolutionary cell biology. J Mol Biol 2020;432:1861–1879.
[251] Kamdar S, Ghosh D, Lee W, Tatulea-Codrean M, Kim Y, Ghosh S, et al. Multiflagellarity leads to the size-independent swimming speed of peritrichous bacteria. Proc Natl Acad Sci USA 2023;120:e2310952120.
[252] Cael BB, Cavan EL, Britten GL. Reconciling the size-dependence of marine particle sinking speed. Geophys Res Lett 2021;48: e2020GL091771.
[253] Simonoff JS. Smoothing methods in statistics. Springer; 2012.
[254] Velho Rodrigues MF, Lisicki M, Lauga E. The bank of swimming organisms at the micron scale (BOSO-Micro). PLoS One 2021;16: e0252291.